\documentclass{article}

\PassOptionsToPackage{numbers, compress}{natbib}

\usepackage[preprint]{neurips_data_2024}

\usepackage[utf8]{inputenc} 
\usepackage[T1]{fontenc}    
\usepackage{hyperref}       
\usepackage{url}            
\usepackage{booktabs}       
\usepackage{amsfonts}       
\usepackage{nicefrac}       
\usepackage{microtype}      
\usepackage[table]{xcolor}         

\usepackage{amsmath}
\usepackage{amssymb}
\usepackage{mathtools}
\usepackage{amsthm}
\usepackage{multirow}
\usepackage{wrapfig}
\usepackage{lipsum}
\usepackage{mwe}
\usepackage[capitalize,noabbrev]{cleveref}
\usepackage{graphicx}
\usepackage{subfigure} 
\usepackage{float}
\usepackage{algorithm} 
\usepackage{algorithmic}
\usepackage{textcomp} 
\usepackage{listings} 
\usepackage{enumitem}
\usepackage{bbding}
\usepackage{tipa}
\usepackage{xcolor}
\usepackage[all]{hypcap}

\usepackage{array}     
\usepackage{tabularx}
\makeatletter
\renewcommand{\@fnsymbol}[1]{%
  \ifcase#1\or
  \dagger\or 
  *\or       
  \ddagger\or
  \mathsection\or
  \mathparagraph\or
  \|\or
  **\or
  \dagger\dagger\or
  \ddagger\ddagger\else\@ctrerr\fi}
\makeatother

\title{ASAudio: A Survey of  Advanced Spatial Audio Research}

\author{%
\normalsize
Zhiyuan Zhu\thanks{Equal contribution}\quad
Yu Zhang\footnotemark[1]\quad
Wenxiang Guo\footnotemark[1]\quad
Changhao Pan\footnotemark[1]\quad
Zhou Zhao\thanks{Corresponding Author}\quad\\
 Zhejiang University\\
\normalsize
\texttt{\{schmittzhu, yuzhang34, zhaozhou\}@zju.edu.cn}\vspace{-1em}
}

\begin{document}

\maketitle

\begin{abstract}
  With the rapid development of spatial audio technologies today, applications in AR, VR, and other scenarios have garnered extensive attention. 
  Unlike traditional mono sound, spatial audio offers a more realistic and immersive auditory experience. 
  Despite notable progress in the field, there remains a lack of comprehensive surveys that systematically organize and analyze these methods and their underlying technologies.
  In this paper, we provide a comprehensive overview of spatial audio and systematically review recent literature in the area. 
  To address this, we chronologically outlining existing work related to spatial audio and categorize these studies based on input-output representations, as well as generation and understanding tasks, 
  thereby summarizing various research aspects of spatial audio. 
  In addition, we review related datasets, evaluation metrics, and benchmarks, offering insights from both training and evaluation perspectives. 
  Related materials are available at \url{https://github.com/dieKarotte/ASAudio}.
\end{abstract}

\section{Introduction}
\label{sec: intro}

Spatial audio is an emerging audio technology designed to provide users with a more immersive and realistic auditory experience\cite{poeschl2013integration}. 
By simulating the way human ears receive sound in three-dimensional space, spatial audio processes the direction, distance, and spatial position of sound events, enabling users to perceive the origin and location of sounds and thereby recreating the natural hearing experience.
Spatial audio typically encompasses positional information such as distance and direction as well as dynamic attributes like velocity and movement direction, in addition to the traditional acoustic properties of sound sources. 
Moreover, it incorporates contextual cues related to the acoustic environment, recognizing that the same sound source will behave differently in a highly absorptive recording studio versus an open space with pronounced reverberation. 
Compared to traditional stereo or surround sound systems, the enriched information in spatial audio results in an enhanced sense of enclosure, improved sound diffusion, and superior localization and directional cues.
It represents a significant advancement in audio technology, with applications already being adopted in industries such as film, television, virtual and augmented reality (VR/AR), video games \cite{broderick2018importance, murphy2011spatial}, music, and museums, all aiming to immerse audiences and elevate the auditory experience.

The evolution of spatial audio can be traced back to the advent of stereo sound. Initially, audio signals evolved from monophonic to stereo, where the use of two channels creates a sense of left-right directionality, introducing a preliminary spatial dimension. 
With technological advancements, surround sound systems expand upon stereo by incorporating additional channels further enhancing spatial perception. 
By strategically positioning speakers around the listener (including in front, at the sides, behind), sound can emanate from a 360-degree horizontal plane, creating a more immersive auditory environment with precise spatial localization and a stronger sense of envelopment.
In natural environments, sound is inherently three-dimensional; however, conventional stereo and surround sound systems are limited to two-dimensional localization, lacking the capacity to convey height and depth. 
Human auditory perception relies on subtle differences between the ears, such as time, intensity, phase, and frequency disparities, to determine the direction and distance of sounds. 
Spatial audio technology replicates these cues to achieve a three-dimensional sound experience, allowing listeners to perceive height, depth, and distance, and thereby creating an immersive auditory scene that faithfully reproduces the acoustic environment as experienced by the human ear.

\begin{figure}[htbp]
  \centering
  \includegraphics[width=\linewidth]{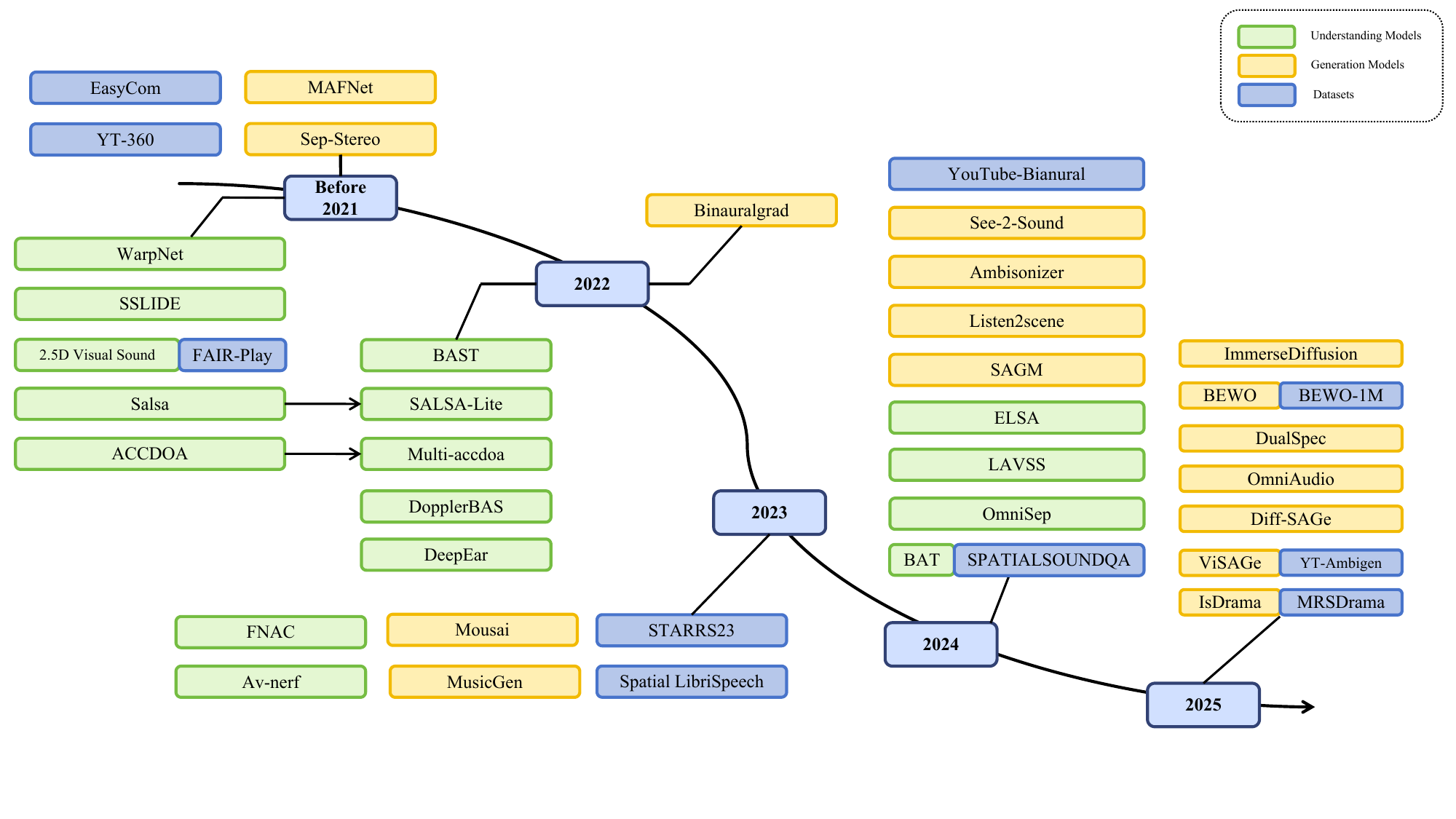}
  \caption{A timeline of recent spatial audio models \& datasets in recent years. The timeline is established
  mainly according to the release date of the technical paper for
  each model. We mark the understanding models in green and the generation models in yellow, while datasets are marked in blue.}
  \label{fig:history}
\end{figure}

As illustrated in Figure \ref{fig:history}, we chronologically summarize the major developments in spatial audio models and datasets. 
These works can be categorized into two groups, corresponding to spatial audio understanding and spatial audio generation tasks.
Spatial audio understanding encompasses sound event detection, source localization, audio separation, and scene acoustics consistency. 
With the growth of available datasets and advances in model design, the diversity of understanding tasks and architectures has steadily increased.
Spatial audio generation covers methods that synthesize multichannel or ambisonic audio from monophonic or multimodal inputs. Over time, the underlying architectures have evolved from early convolutional neural networks (CNNs) to modern Transformer-based, diffusion-based, and flow-matching models, as well as various hybrid designs.
Early spatial audio datasets are largely derived by spatializing monaural recordings or using simulated data. 
More recently, the advent of diverse recording setups and text-audio or video-audio datasets has enriched multimodal spatial audio datasets, many of which now combine multiple modalities.

Prior to 2021, research on spatial audio understanding dominated, extending traditional SELD tasks to spatial audio contexts and audio separation. Representative works include SSLIDE \cite{wu2021sslide}, WarpNet \cite{richard2021neural}, Salsa \cite{nguyen2022salsa}, and ACCDOA \cite{shimada2021accdoa}, as well as the pioneering 2.5D Visual Sound \cite{gao20192} with its own benchmark dataset.
Most of these early models employed CNN architectures (e.g., Sep-Stereo \cite{zhou2020sep}), with a few using Transformers for generation (e.g., MAFNet \cite{zhang2021multi}). 
Datasets from this period, such as EasyCom \cite{donley2021easycom} and YT-360 \cite{morgado2020learning}, are limited in scale and diversity.
From 2022 to 2025, the rapid progress of generative models and the availability of multimodal datasets shift focus toward spatial audio generation. 
Diffusion-based models like ImmerseDiffusion \cite{heydari2025immersediffusion} and flow-matching approaches such as Diff-SAGe \cite{kushwaha2025diff} substantially improve the perceptual quality and spatial realism of binaural, multichannel, and FOA audio. 
Many generation and understanding frameworks introduce their own datasets during this period, including SPATIALSOUNDQA for BAT \cite{zheng2024bat}, MRSDrama for ISDrama \cite{zhang2025isdrama}, YT-Ambigen for ViSAGe \cite{kim2025visage}, and BEWO-1M for BEWO \cite{sun2024both}, thus providing valuable resources for subsequent studies.
Contemporary understanding models have also adopted newer architectures: LAVSS \cite{ye2024lavss} employs attention mechanisms for spatial audio separation, while BAT \cite{zheng2024bat} integrates large language models to interpret spatial audio events and locations from natural language queries.

In recent years, spatial audio has also attracted significant attention and investment from leading technology companies and startups worldwide. 
Apple is among the first to integrate spatial audio into consumer products such as AirPods and Vision Pro\cite{chen2025spatial}, leveraging proprietary head-related transfer function (HRTF) algorithms and sensor-based dynamic head-tracking to optimize spatial rendering and deliver a highly immersive listening experience . 
Apple Music has also launched a spatial audio streaming service, supporting formats like Dolby Atmos\cite{wuolio2023potential} . 
Google\cite{google2025article} has added native support for spatial audio in Android, releasing a framework for Ambisonics and object-based audio rendering that has accelerated adoption on mobile platforms .
Meta has made spatial audio a key pillar of its metaverse strategy, embedding it in Oculus VR and other virtual reality devices to enhance realism in social and immersive content.

When constructing spatial audio structure, we have identified four major technical challenges based on different application scenarios and task requirements. 
First is the design of spatial audio representations. 
For the input modality, it is essential to consider how to represent the spatial information corresponding to sound events, ensuring consistency among various multimodal inputs; 
While for the output modality, the focus is on the reconstruction and generation of spatial audio.
The second challenge lies in understanding the spatial information of audio events, such as location, direction, and distance. 
Third is the generation of spatial audio with effective spatial cues, which involves exploring different model architectures, generation strategies, and training methods. 
The fourth challenge pertains to data: existing spatial audio datasets are limited and lack standardized evaluation methods and metrics.

\begin{figure}[htbp]
    \centering
    \includegraphics[
        trim={0cm 8cm 0cm 5.3cm}, 
        clip, 
        width=0.95\textwidth, 
        keepaspectratio
    ]{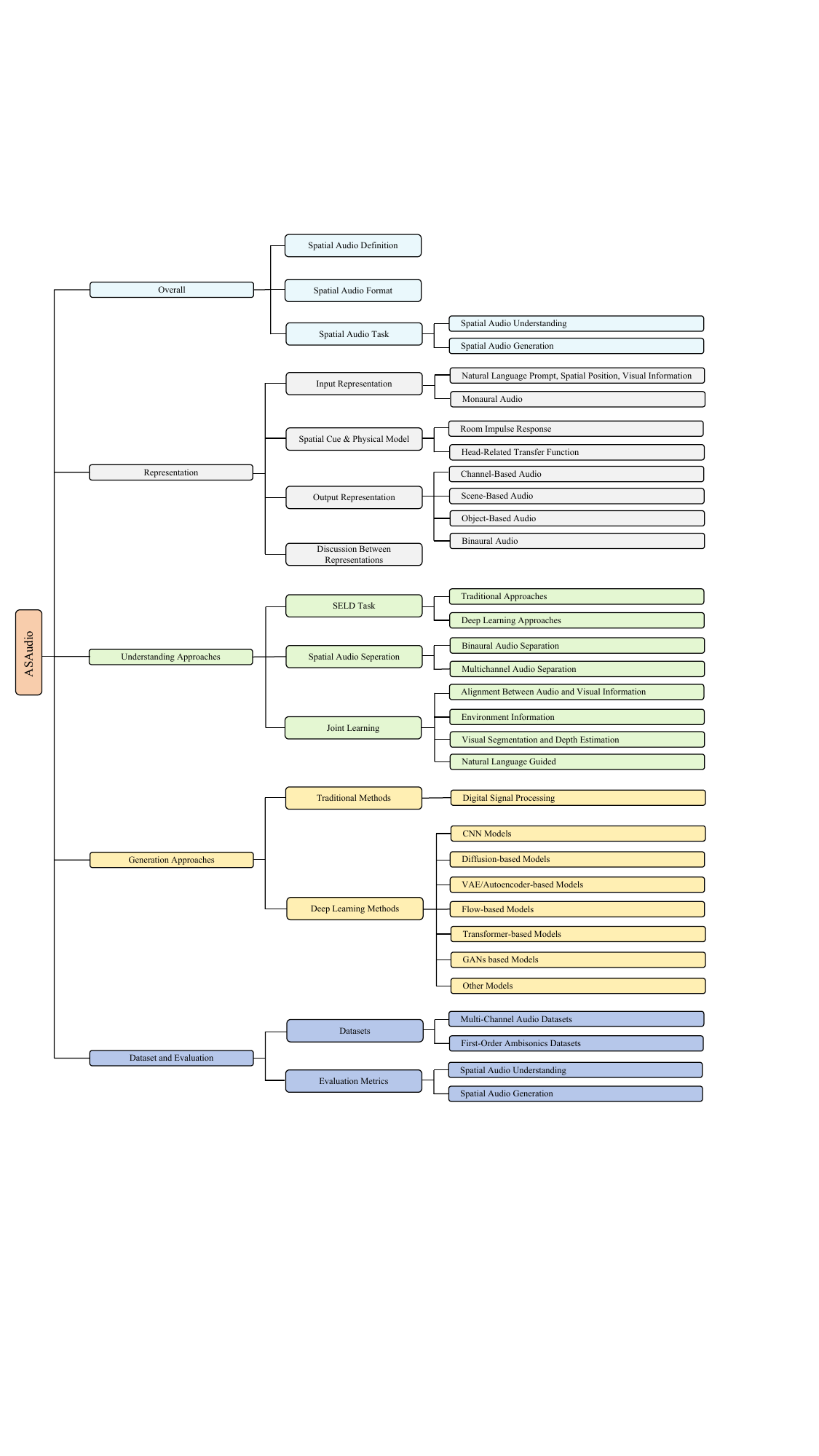}
    \caption{A general overview about the structure of ASAudio.}
    \label{fig:structure}
\end{figure}

Given these considerations, our study sequentially addresses these technical challenges. 
First, we discuss spatial audio representation methods. In Section 3, we introduce and compare the input-output modalities of the primary spatial audio tasks. 
Next, Section 4 focuses on spatial audio understanding tasks, including event detection, localization, spatial audio separation, spatial acoustic scene consistency, and several joint learning approaches. This section primarily analyzes how to interpret and understand the manifestation of spatial audio based on its acoustic features. 
Section 5 then presents spatial audio generation tasks, covering traditional audio spatialization techniques and spatial audio generation tasks based on various deep learning model architectures. 
Finally, Section 6 summarizes the available spatial audio datasets and evaluation methods, including various types of datasets, construction methods, and evaluation metrics.
We provide an overview of the structure of this paper in Figure \ref{fig:structure}.

\section{Overall}
\label{sec: over}

\subsection{Spatial Audio}
Spatial audio is an advanced audio technology that aims to recreate a three-dimensional auditory experience for listeners, going beyond the limitations of traditional stereo or surround sound systems. 
By simulating how human ears perceive sound from different directions, distances, and heights in real environments, spatial audio enables users to accurately sense the position, movement, and depth of sound sources. 
This is achieved through the processing of spatial cues such as interaural time difference (ITD), interaural level difference (ILD), and head-related transfer functions (HRTF), as well as the integration of dynamic information like head movement and environmental acoustics.

Spatial audio technology is widely applied in various fields, including virtual reality (VR), augmented reality (AR), gaming, film, music, and teleconferencing, providing a more immersive and realistic listening experience. 
With the rapid development of hardware devices and rendering algorithms, spatial audio has become a key component in enhancing user immersion and interaction in digital content and smart devices.

\subsection{Spatial Audio Format}
Spatial audio can be conveyed in three principal formats. 
First, channel-based systems such as 5.1 or 7.1 surround sound assign discrete audio channels to predefined speaker locations.
During mixing, each channel is mapped to its corresponding speaker, and during playback, the signal intended for each channel is routed to the appropriate loudspeaker.
Second, scene-based formats like first-order Ambisonics (FOA) employ spherical harmonic decomposition to represent the entire three-dimensional sound field. 
These signals are then decoded through a higher-order Ambisonics (HOA) decoder into multichannel outputs for traditional speaker arrays. 
Third, object-based approaches—exemplified by Dolby Atmos—model each sound source as an independent object with associated metadata specifying its position and movement.
At playback, the renderer dynamically projects these objects to available speakers or binaural renderers, enabling true three-dimensional audio placement.

\subsection{Spatial Audio Tasks}
The main research tasks in the field of spatial audio can be broadly divided into two categories: spatial audio understanding and spatial audio generation. 
Spatial audio understanding focuses on analyzing and interpreting spatial audio signals to extract information about sound events, source locations, and the acoustic environment. 
In contrast, spatial audio generation aims to synthesize realistic three-dimensional audio experiences from limited or monaural inputs, often leveraging multimodal cues. 
These two directions complement each other and together drive the advancement of immersive audio technologies.

Currently, spatial audio is often integrated with multiple modalities, such as video input for scene information, natural language descriptions of sound source locations, and monaural audio timbre prompts. 
By leveraging this multimodal information, spatial audio can be better understood and generated.

\subsection{Spatial Audio Understanding}
Spatial audio understanding refers to a set of tasks focused on analyzing and interpreting spatial audio signals to extract meaningful information about the acoustic scene. 
The main objectives include detecting and classifying sound events, localizing sound sources in three-dimensional space, separating overlapping sources, and characterizing the overall acoustic environment. 
These tasks are fundamental for applications such as spatial sound event detection, sound source localization, audio-visual scene analysis, and intelligent human-computer interaction.

Traditional approaches to spatial audio understanding often rely on signal processing techniques such as beamforming, time-delay estimation, and spatial filtering to extract spatial features and cues like ITD, ILD and HRTF.
While effective in controlled environments, these methods may struggle with complex or dynamic real-world scenes.

With the advancement of deep learning, modern spatial audio understanding systems increasingly employ neural networks, including convolutional neural networks (CNNs), recurrent neural networks (RNNs), and Transformers, to learn robust representations from raw spatial audio signals or their time-frequency features. 
These models can be further enhanced by incorporating multimodal information, such as video frames or textual descriptions, enabling richer scene understanding and cross-modal reasoning.
Recent research also explores the integration of large language models (LLMs) and attention mechanisms to interpret spatial audio in the context of natural language queries or to focus on salient sound events and locations.

Overall, spatial audio understanding is a rapidly evolving field that combines signal processing, machine learning, and multimodal fusion to enable intelligent perception and interaction in immersive environments.

\subsection{Spatial Audio Generation}
Spatial audio generation refers to the task of synthesizing audio signals that provide listeners with a realistic three-dimensional auditory experience, often from monaural or limited-channel inputs. 
The goal is to generate spatial cues—such as direction, distance, and movement of sound sources—so that the resulting audio can be perceived as coming from specific locations in space, thereby enhancing immersion and realism.

The core challenge in spatial audio generation lies in accurately modeling and reconstructing the spatial characteristics of sound, including ITD, ILD and HRTF cues. 
This task is crucial for applications in virtual reality (VR), augmented reality (AR), gaming, film, and music production, where immersive audio is essential.

Early approaches to spatial audio generation primarily relied on traditional digital signal processing (DSP) techniques, such as applying fixed delays, filters, or measured HRTFs to monaural signals to simulate spatial effects. 
While these methods are intuitive and computationally efficient, they often lack perceptual realism, especially in dynamic or complex acoustic environments.

With the advent of deep learning, spatial audio generation has seen significant advancements. 
Modern methods leverage neural networks such as convolutional neural networks (CNNs), U-Net architectures, variational autoencoders (VAEs), diffusion models, Transformers, and flow-matching frameworks to learn complex mappings from input modalities (e.g., monaural audio, video, text, or spatial metadata) to spatialized audio outputs. 
These models can capture intricate relationships between audio content, visual cues, and spatial context, enabling end-to-end generation of high-quality binaural, multichannel, or ambisonic audio.

Recent research also explores multimodal spatial audio generation, where models are conditioned on additional inputs such as video frames, scene geometry, or textual descriptions to further enhance spatial accuracy and realism. 
For example, visually guided models use video information to infer the position and movement of sound sources, while text-conditioned models generate spatial audio based on semantic scene descriptions.

Overall, spatial audio generation is a rapidly evolving field that integrates signal processing, machine learning, and multimodal data fusion to push the boundaries of immersive.
\section{Representations of Spatial Audio}
\label{sec: repre}
Spatial audio representations are diverse. This section introduces the input and output representations of spatial audio tasks, followed by a discussion and comparison of these representations.

\subsection{Inputs Representations}
The input representations for spatial audio tasks vary widely.
Given that spatial audio encompasses semantic, acoustic, and spatial information, the input representations are designed to capture these aspects.
For the acoustic component, monaural audio prompts are commonly used as input, providing acoustic features such as timbre.
For the semantic component, text prompts are typically used, offering semantic information about the audio.
For the spatial component, spatial position prompts can take various forms, including natural language text descriptions, visual information in the form of images or videos, or spatial position coordinates.
These input representations can be used individually or in combination, often with different combinations tailored to specific tasks.

\begin{figure}[htbp]
  \centering
  \label{fig:input_format}
  \includegraphics[
        trim={0cm 0.2cm 0cm 0.3cm}, 
        clip, 
        width=\textwidth]{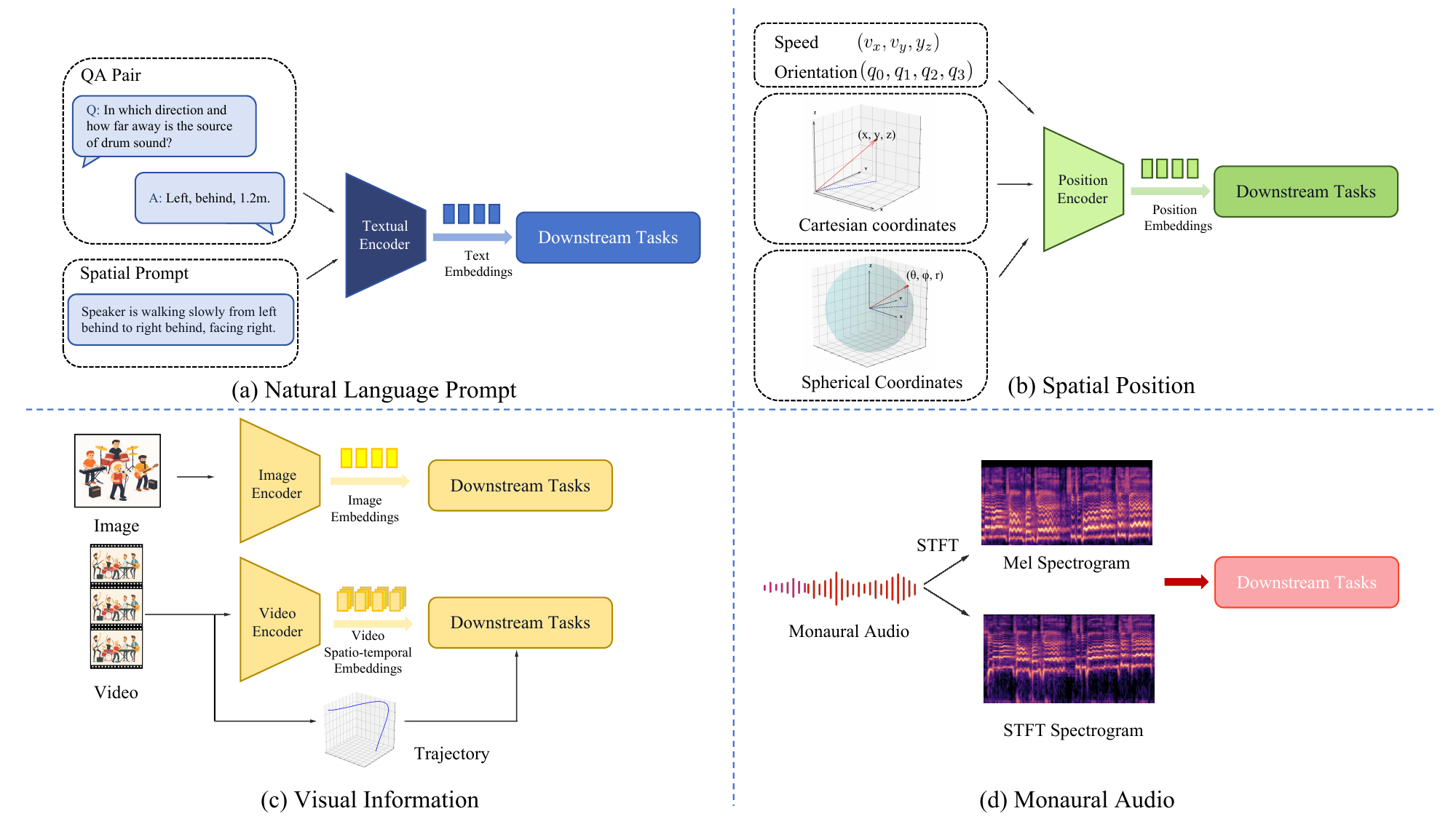}
  \caption{An overview of input representations of spatial audio and their primary processing methods.}
\end{figure}

We illustrate the various input representations and their primary processing methods in Figure \ref{fig:input_format}.
The indirect formats including text, visual, and spatial position are processed by the encoder to extract the corresponding feature embeddings, while mono audio can be directly used by end-to-end models or used after short-time Fourier transform (STFT) to obtain the spectrogram.
Also the video visual input contains time domain information, where trajectories and spatial-temporal features can be extracted.

\subsubsection{Natrual Language Prompts}
Natural language prompts serve as a powerful and intuitive way to specify the semantic content and, in some cases, the spatial characteristics of an audio scene. 
In generation tasks, text can describe the event to be synthesized (e.g., "a bird chirping on the left") \cite{kreuk2022audiogen, liu2023audioldm}. 
In understanding tasks, text can form a query about the audio scene (e.g., "Is the dog barking behind me?"), requiring the model to reason about both the content and spatial arrangement. 
BAT \cite{zheng2024bat} is an example of a model that uses large language models. It handles spatial audio tasks by incorporating natural language question-answer pairs related to sound event detection, direction and distance estimation, and spatial reasoning.
By training in stages, it learns to handle different types of question-answer pairs, gradually improving its performance in understanding the spatial information embedded in natural language.
The natural language modality bridges the gap between high-level human intent and low-level audio signal processing.

\subsubsection{Spatial Position}
Explicit spatial position data, such as Cartesian coordinates (x, y, z) or spherical coordinates (azimuth, elevation, distance), provides direct and unambiguous guidance for spatial audio tasks. 
In generation, this information is used to place a synthesized sound source at a precise location in the 3D space. 
In understanding, it serves as the ground truth for training and evaluating sound source localization models. 
Recent works \cite{liu2022dopplerbas, zhang2025isdrama} have also incorporated additional spatial position information, such as the radial velocity and orientation of sound sources, to enhance the dynamic characteristics of spatial audio by simulating the Doppler effect produced during real sound source movement.

\subsubsection{Visual Information}
Visual information, from static images to dynamic videos, is one of the richest sources of spatial and semantic context. 
The strong correlation between what we see and what we hear makes vision an invaluable input for spatial audio tasks. 
The visual-acoustic matching task \cite{chen2022visual} is crucial in mono audio generation, which also holds significant importance in spatial audio generation.
In audio-visual source separation and localization, the visual presence of an object provides a strong cue for isolating and locating its corresponding sound \cite{zhao2018sound, ye2024lavss, zhou2018visual}. 
While in generation, \cite{gan2019self, gao20192} propose methods that utilize visual cues to guide the synthesis of spatial audio from monaural inputs with a U-Net architecture, effectively decoding the audio in a spatially aware manner.

\subsubsection{Monoaural Audio}
Monoaural audio serves as the foundational acoustic content in many spatial audio generation tasks.
It provides the core acoustic characteristics, such as the timbre of a specific instrument or the phonetic features of speech, which are then spatialized by the generation model to determine "where" the sound is located.
In two-stage spatial audio generation systems, the first stage often processes monaural audio streams and "upmixes" them into multichannel or binaural formats, while considering additional spatial position inputs such as visual or positional data.
This approach allows the model to focus on the spatial characteristics of the sound while leveraging the rich acoustic information contained in the monaural input.

\subsection{Spatial Cues and Physical Modeling}
A fundamental aspect of spatial audio is the accurate modeling of how sound propagates and is perceived in three-dimensional space. 
Two key concepts in this context are the room impulse response (RIR) and the head-related transfer function (HRTF).

\subsubsection{Room Impulse Response (RIR)}
Under the assumption that the acoustic environment is a Linear and Time-Invariant (LTI) system, the Room Impulse Response (RIR) completely characterizes all acoustic paths between a sound source and a receiver. 
The RIR can be decomposed into three main components: the direct sound, early reflections, and late reverberation. 
Once the RIR is obtained, any anechoic signal can be precisely predicted for that specific room and location through convolution. 
Therefore, the RIR serves as the bridge between virtual acoustics and the real world.

The traditional method for acquiring an RIR involves precise measurements in a controlled environment. 
However, measuring a large number of source-receiver pairs in a space is exceedingly tedious and time-consuming, which constitutes a major bottleneck for the practical application of RIRs. 
To overcome this challenge, the research community has developed various methods. 
One of the ultimate goals is to actively mitigate or alter the adverse effects of room acoustics, such as applying dereverberation to enhance speech clarity or creating realistic virtual acoustic environments in Virtual Reality (VR) and Augmented Reality (AR).

\begin{itemize}
    \item \textbf{RIR Estimation from Visual Information} \quad A leading research direction is the "blind" estimation of RIRs using visual information, thereby avoiding complex acoustic measurements. One study proposed a system that can estimate room geometry and acoustic properties from 360° panoramic images to generate RIRs for VR/AR. Another work employed a multi-modal learning approach to jointly estimate the RIR from a reverberant speech signal and visual cues of its environment. Building on this, \cite{majumder2022few} utilized a Transformer model to infer the RIR of an entire space from just a few sparse images and their corresponding echoes, enabling few-shot generalization to new environments.
    \item \textbf{RIR Simulation and Application} \quad Besides measurement or estimation, RIRs can also be generated through simulation and used as a core tool in various audio tasks. The classic image-source method provides an efficient way to compute RIRs for simple rectangular rooms. Modern software libraries like SpatialScaper \cite{roman2024spatial} use RIR simulation to generate complex soundscapes for training machine learning models for tasks like Sound Event Localization and Detection (SELD). In terms of application, the work of \cite{ahn2023novel} demonstrates that using RIRs generated from 3D room models can significantly improve a neural network's performance on tasks such as sound source separation and dereverberation. Similarly, RIRs are also simulated to generate data for various other purposes \cite{jeub2009binaural, vacher2014sweet, mittag2017dataset, di2021dechorate, grondin2020bird}. The closely related Head-Related Impulse Response (HRIR) is also used to generate training data for binaural audio \cite{xu2021visually}.
    \item \textbf{RIR Measurement and Analysis for Specific Scenarios} \quad To support more complex applications, researchers have conducted precise RIR measurements for specific scenarios. For instance, to evaluate high-resolution sound field analysis methods, \cite{koyama2021meshrir} performed RIR measurements on a dense grid of points, while \cite{ratnarajah2022mesh2ir} developed a neural network-based method to predict impulse responses in real-time from a given 3D scene mesh. Another work \cite{di2021dechorate} provided RIRs with a focus on being echo-aware to support research in echo-related signal processing. To address dynamic scenes, researchers have captured RIRs of moving sources \cite{politis2020dataset} to advance SELD technology. Furthermore, to achieve six-degrees-of-freedom (6DoF) immersive experiences, \cite{mckenzie2021acoustic,mckenzie2021dataset} measured spatial RIRs (SRIRs) in a room with variable acoustics.
    \item \textbf{Perceptual Evaluation based on RIR} \quad The ultimate goal of all RIR-related technology is to achieve auditory realism, making perceptual evaluation essential. For example, \cite{brinkmann2017authenticity} utilized binaural RIRs (BRIRs) measured in different reverberant environments to conduct subjective listening tests and perceptual attribute assessments on the authenticity of dynamic binaural synthesis. Likewise, \cite{di2021dechorate} provides a tool for benchmarking recent methods in a variety of echo-aware tasks, including RIR estimation and room geometry estimation.
\end{itemize}

\subsubsection{Head-Related Transfer Function (HRTF)}
The Head-Related Transfer Function (HRTF) is a subject-specific acoustic filter that describes how sound is altered by an individual's anatomy—primarily the head, torso, and pinnae—before reaching the eardrums. It encodes the essential cues for 3D sound perception: binaural cues (Interaural Time and Level Differences, ITD/ILD) for horizontal localization, and monaural spectral cues for elevation and front-back discrimination. This filtering process can be modeled as a Linear Time-Invariant (LTI) system.

This inherent anatomical dependency makes HRTF personalization crucial for immersive audio. Using non-personalized or generic HRTFs leads to significant perceptual errors, including poor elevation judgment, front-back confusion, and a lack of externalization known as In-Head Localization (IHL). These distortions severely degrade the realism of the auditory experience. Consequently, a primary goal of spatial audio research is to develop personalization methods that overcome these issues by effectively balancing accuracy, cost, and convenience.

To address this challenge, the research community has developed various HRTF personalization methods, which present different trade-offs between accuracy, cost, and convenience.

\paragraph{Customization based on Anthropometry}
This category of methods attempts to predict a user's HRTF by measuring their physiological characteristics. For instance, \cite{warnecke2022hrtf} studied HRTF personalization from a morphological perspective, confirming that similarity in ear shape correlates with better localization performance. Subsequent work has focused on using neural networks to predict critical spectral features for vertical localization, such as the N1 notch, from ear measurements \cite{arbel2024hrtf}. Further advancing this, some research \cite{zhao2022magnitude} has combined anthropometric data with ear images, using Convolutional Neural Networks (CNNs) to predict the complete HRTF.

\paragraph{Database-driven and Perception-guided Personalization}
These methods aim to select or recommend an optimal matching HRTF for a user from a large database. To achieve this, one study \cite{lee2023global} utilized machine learning to construct an HRTF error metric consistent with human auditory perception. Another study \cite{marggraf2024hrtf} directly built a recommendation model by predicting the timbral coloration distortion that different HRTFs might cause.

\paragraph{Spatial Upsampling from Sparse Data}
This is currently the most mainstream research direction, which involves using a small number of precise acoustic measurements to generate a complete, high-resolution HRTF through interpolation by deep learning models.

We will summarize and discuss the current state of HRTF personalization research from the following aspects:
\begin{itemize}
\item \textbf{Basic Deep Learning Architectures} \quad
Early work successfully applied mainstream deep learning models to this task. 
For example, \cite{jiang2023modeling} treated the HRTF as a 2D image and used a CNN for super-resolution reconstruction. 
Another work \cite{ito2022head} proposed using an autoencoder conditioned on source position for efficient HRTF interpolation from sparse measurements. 
To generate more realistic details, \cite{hogg2024hrtf} introduced Generative Adversarial Networks (GANs). 
More cutting-edge research \cite{ma2023spatial} has employed VQ-VAE and Transformer architectures. 
To address the issue of model generalization across different datasets, \cite{zhang2023hrtf} proposed a novel architecture featuring a hypernetwork.
\item \textbf{Fusing Physical and Geometric Priors} \quad
To enhance performance, researchers began to integrate physical or geometric prior knowledge into their models. 
\cite{chen2023head} proposed a Spherical CNN to better handle the inherent spherical geometry of HRTF data. 
Another work \cite{thuillier2024hrtf} utilized a meta-learning framework to efficiently learn how to correct an initial HRTF estimate with a few samples.
\item \textbf{The Neural Field Paradigm} \quad
Neural Fields have elevated the representation of HRTFs from discrete data points to a continuous function. 
A pioneering work \cite{zhang2023hrtf} proposed the HRTF-Field, which uses a neural network to represent the entire HRTF, allowing for queries from any spatial direction. 
Subsequently, \cite{masuyama2024niirf} introduced NIIRF, which enables the neural field to directly predict the coefficients of efficient IIR filters, resulting in a more compact and physically motivated model.
\end{itemize}
HRTF personalization technology is undergoing a profound transformation from physical modeling to data-driven approaches, from pursuing signal accuracy to optimizing perceptual experience, and from discrete representations to continuous functions. 
Future research may fuse the advantages of different methods, for example, by combining anthropometry \cite{warnecke2022hrtf} with sparse upsampling \cite{jiang2023modeling}. 
The ultimate goal is to deploy these powerful models on consumer-grade devices, finally achieving seamless, high-fidelity personalized spatial audio experiences for everyone.

\subsection{Outputs Representations}
The output representations for spatial audio tasks are diverse, typically including binaural audio, stereo, multichannel audio, and Ambisonics. 
We will introduce the output representations in three categories: channel-based audio, scene-based audio, and object-based audio and will separately discuss binaural audio rendering technology.

To achieve the goal of an immersive auditory experience, various audio representation formats have been developed for encoding and reproducing spatial information. 
We will systematically analyze the output representations of spatial audio by dividing them into three major paradigms: Channel-Based Audio, Scene-Based Audio, and Object-Based Audio. 
The channel-based format is the most traditional approach, directly mapping audio signals to predefined loudspeaker positions. 
The scene-based format aims to physically reconstruct the complete sound field within a specific region. 
In contrast, the object-based format represents the most flexible and abstract method, as it completely decouples audio content from the physical playback configuration. 
These different audio formats are ultimately decoded and rendered through headphones, loudspeaker arrays, or other audio devices to provide the final auditory experience.

\subsubsection{Channel-Based Audio}
Channel-based audio is the most traditional and widely known representation method in the field of spatial audio. 
Its core principle is the direct and fixed binding of audio signals to specific channels of a physical playback device. 
It leverages psychoacoustic principles to "deceive" the human auditory system, thereby creating a sense of space between a limited number of loudspeakers.
The fundamental concept of this paradigm is that audio signals are pre-mixed into multiple independent channels. 
Each channel corresponds explicitly to a predetermined loudspeaker position. For example, a traditional stereo system consists of left and right channels, which are delivered to the left and right loudspeakers, respectively. 
More complex surround sound systems, such as 5.1 or 7.1 channel systems, include center, front-left, front-right, surround-left, surround-right, rear-surround, and a Low-Frequency Effects (LFE) channel. 
Each of these corresponds to a specific loudspeaker position in the listening environment. 
The method of encoding spatial information is implicit. 
The spatial position of a sound is not recorded as independent data but is instead implied through level and time differences among the various channels. 
The psychoacoustic basis for this process is "summing localization"". 
Experiments have shown that when a pair of loudspeakers plays coherent signals with a specific level or time difference, the listener perceives a single "phantom image" between them. 
For instance, by controlling the amplitude ratio of a monophonic source across two or more channels, a technique known as Amplitude Panning, one can create the perception of a virtual source's position. 
At low frequencies ($f \le 0.7kHz$), the perceived azimuth $\theta_{I}$ of the phantom image follows the sine law with respect to the amplitude $E_{L}$ and $E_{R}$ of the left and right loudspeakers, and the half-angle $\theta_{0}$ between them:
$sin~\theta_{I}=\dfrac{E_{L}-E_{R}}{E_{L}+E_{R}}sin~\theta_{0}.$
In addition to amplitude panning, more advanced techniques such as Vector Base Amplitude Panning (VBAP) and Distance-Based Amplitude Panning (DBAP) enable more precise source localization in complex multi-loudspeaker arrays.

The development of channel-based systems has evolved from two-channel stereo (1930s) and quadraphonic sound (1970s) to the widely used 5.1 and 7.1 channel systems of today. 
The failure of quadraphonic systems was due to their coarse and uniform sampling of the horizontal sound field, which neglected the non-uniformity of auditory perception and resulted in unstable sound images. 
In contrast, the success of 5.1 channel sound, particularly in applications with accompanying video, is attributed to its non-uniform design philosophy. 
It allocates more resources to accurately reproduce the frontal sound field (three front channels), while the side and rear surround channels are primarily used to create environmental ambiance. 
This approach better aligns with human attentional focus in audiovisual scenarios.
As technology has advanced, the number of channels has increased further, leading to systems that include height information, such as 9.1, 10.2, and even 22.2 channel systems. 
These systems use layered loudspeaker layouts to create a stronger sense of three-dimensional space.

\subsubsection{Scene-Based Audio}
In contrast to channel-based methods, the scene-based audio paradigm, also known as sound field reproduction, is not limited to creating auditory illusions at specific points. 
Instead, it aims to completely capture and physically reproduce the entire sound field within a defined spatial region. 
Its core philosophy is to use physics-based models to analyze and synthesize the complete acoustic wave field in a specific area. 
In theory, this allows listeners to move freely within this reproduced region and experience stable, consistent spatial audio, thereby overcoming the limitation of a single "sweet spot".
This paradigm primarily involves two key technologies: Ambisonics and Wave-Field Synthesis (WFS).
The mathematical foundation of Ambisonics is the decomposition of a 3D sound field using Spherical Harmonics. 
In a source-free region, the sound pressure $P(x,\omega)$ at any point can be represented by an expansion of spherical harmonic basis functions:
$ P(x, \omega) = \sum_{n=0}^{N} \sum_{m=-n}^{n} \alpha_{n}^{m}(\omega) j_{n}(kr) Y_{n}^{m}(\hat{x}) $
where $\alpha_{n}^{m}(\omega)$ are the sound field coefficients, which contain all directional information of the sound field; $j_{n}(kr)$ is the spherical Bessel function, describing the radial propagation characteristics of sound waves; and $Y_{n}^{m}(\hat{x})$ are the spherical harmonic functions, describing the angular distribution of the sound field.
The term $N$ represents the order of the system, which determines the spatial resolution of the sound field description. 
When $N=1$, the system is referred to as first-order Ambisonics (FOA), commonly known as B-format. 
When $N \ge 2$, it is high-order Ambisonics (HOA).
A higher order yields greater spatial resolution but also dramatically increases the required data volume and computational complexity. 

An Ambisonic sound field is typically recorded using specialized microphone arrays (such as tetrahedral or spherical arrays). 
During playback, these signals are converted by a decoder into driving signals suitable for any given loudspeaker layout.
Wave-Field Synthesis (WFS) is another significant sound field reproduction technique, based on the Kirchhoff-Helmholtz Integral from acoustics. 
This principle states that the sound field inside a source-free region can be perfectly reconstructed by a continuous distribution of secondary sources on the boundary of that region. 
In practice, this ideal continuous distribution is approximated by an array of numerous, densely packed, and independently driven loudspeakers. 
Unlike the modal decomposition approach of Ambisonics, WFS is more akin to an acoustic form of holography, attempting to project a virtual wavefront into the listening area.

The scene-based paradigm is theoretically superior, with its primary advantage being the creation of a large and stable listening area that allows multiple listeners to move freely. However, both Ambisonics and WFS have extremely high system requirements. 
They demand a large number of loudspeakers, complex real-time digital signal processing capabilities, and high-bandwidth data transmission. 
This has largely limited their adoption in the consumer market, confining their use primarily to scientific research, virtual reality, and professional-grade projects. 
This pursuit of physical accuracy is both the paradigm's greatest strength and its greatest weakness. 
Finding a balance between physical fidelity and engineering practicality remains a central issue in its development. 

\subsubsection{Object-Based Audio}
Object-based audio is the most modern and revolutionary method for spatial audio representation. It achieves unprecedented flexibility and scalability by completely decoupling the audio content from channel allocation. 
Its core innovation is that audio is no longer pre-mixed into fixed channels. 
Instead, the audio content is packaged into individual "audio objects". 
Each object comprises two components: the pure sound material itself (e.g., a line of dialogue or an explosion sound), and a set of descriptive metadata. 
This metadata specifies the object's position, size, motion trajectory, and other acoustic attributes in three-dimensional space.
A fundamental shift in this paradigm is the postponement of the final mixing process. 
The mix is not completed once in the production studio; instead, it is generated in real-time on the end-user's playback device through a process called "rendering". 
The renderer on the playback device reads the metadata and dynamically calculates the necessary loudspeaker signals to position the audio object correctly in space, based on the device's current loudspeaker configuration or headphones. 
For instance, if a helicopter sound is tagged with metadata as "from directly above", the system will direct the signal primarily to any detected overhead speakers. 
In commercial applications, Dolby Atmos, DTS:X, and MPEG-H 3D Audio are prominent examples of the object-based audio paradigm. 
Notably, these modern audio formats typically use a hybrid structure. 
They contain a traditional, channel-based "bed"—for instance, a 7.1 channel layout—to carry ambient sounds and background music. 
Superimposed on this bed can be up to hundreds of independent, dynamic audio objects, which can be positioned with high precision anywhere in three-dimensional space.

In essence, object-based audio represents the highest level of abstraction in spatial audio encoding, shifting the paradigm from producing sound recording to delivering an auditory experience.

\subsubsection{Binaural Audio}
The three aforementioned paradigms are different formats for encoding and distributing spatial audio. In contrast, binaural audio is a crucial rendering technology; 
it is the final form through which all advanced spatial audio formats are delivered to the human ear via headphones.
Binaural audio technology is designed to mimic the human auditory system, reproducing sound specifically for the listener's two ears, typically through headphone playback. 
Its ultimate goal is to accurately reconstruct at the listener's eardrums the sound pressure waves that would be received in a real acoustic environment. 
This process deceives the brain into perceiving a complete three-dimensional sound field. 
The core of this process is the Head-Related Transfer Function (HRTF). HRTF acts as a digital filter that mathematically describes how a sound wave is altered by the complex morphology of the listener's head, torso, and outer ears before reaching the eardrum. 
A pair of HRTFs completely encodes all the key physical cues required for human sound localization, including ITD, ILD, and direction-dependent spectral cues caused by the pinnae.

However, creating a convincing binaural experience requires more than just static HRTF filtering; it must also incorporate dynamic and environmental cues. 
Dynamic cues are implemented through head tracking, where the system tracks the listener's head movements in real-time and updates the applied HRTFs accordingly. 
This effectively resolves front-back confusion and significantly enhances the sense of localization, even when using non-individualized HRTFs.
Environmental cues are realized by simulating room acoustic effects, namely reverberation. 
If only the direct sound is rendered, the sound often appears to be inside the head, an effect known as "in-head localization". 
To achieve "externalization", where the sound is perceived as coming from outside the head, it is necessary to simulate the room's early discrete reflections and late diffuse reverberation, rendering them with their corresponding HRTFs.

Binaural rendering is the common endpoint for personalized audio. 
All three major spatial audio paradigms discussed can be rendered into a two-channel headphone output. Whether it is a channel-based 5.1 signal, a scene-based Ambisonics signal, or an object-based audio stream, all can be convolved with HRTFs corresponding to the appropriate directions. 
The results are then summed into two signals for headphones, providing a high-quality spatial audio experience for the user.

\subsection{Discussion About Representations in Spatial Audio}
\subsubsection{Input Representations Comparison}
The input modalities for spatial audio tasks, natural language, spatial position, visual information, and monaural audio, each offer a unique perspective for the system to perceive, interpret, or generate soundscapes. 
While they can be used independently, their true potential is often realized through synergistic multimodal combinations. 
The choice of input representation is not merely a technical decision but a fundamental architectural one that dictates the system's capabilities, complexity, and the nature of its interaction with the user or environment. 
This section will comparatively analyze these input paradigms, examining their intrinsic properties, task suitability, and the emerging trends in their combined application.

As shown in Table \ref{tab: com_in}, these input representations exhibit a core trade-off between the level of abstraction and control precision. 
Natural language and visual information reside at the highest level of abstraction. 
They are intuitive for humans and well-suited for high-level scene description or content querying. 
However, this intuitiveness introduces challenges of lower control precision and semantic ambiguity, necessitating complex models to bridge the gap between semantics and machine-processable signals. 

Conversely, spatial position coordinates offer the highest control precision, making them ideal for defining precise source trajectories or serving as ground truth for evaluation. 
However, they lack semantic context, and manually specifying complex scenes is a tedious process. 
Monaural audio plays a unique role. Positioned at a low level of abstraction, it does not directly provide spatial control. 
Instead, it serves as the foundational acoustic content for generation tasks, providing core acoustic features such as timbre and pitch. 
It acts as raw material that other modalities spatialize.

Therefore, the selection of an input representation is fundamentally a trade-off between the intuitive, abstract control preferred by humans and the precise, geometric data required by machines, a choice contingent on the specific requirements of the task.

\begin{table}[ht]
\centering
\caption{Comparative Analysis of Spatial Audio Input Representations}
\label{tab: com_in}
\renewcommand{\arraystretch}{1.5}
\renewcommand{\tabularxcolumn}[1]{>{\centering\arraybackslash}m{#1}}

\begin{tabularx}{\textwidth}{
    >{\bfseries}m{2.8cm}
    >{\hsize=1.0\hsize}X  
    >{\hsize=1.0\hsize}X  
    >{\hsize=1.0\hsize}X  
    >{\hsize=0.8\hsize}X  
}
\toprule
\multicolumn{1}{c}{\textbf{Attribute}} &
\multicolumn{1}{c}{\textbf{\begin{tabular}[c]{@{}c@{}}Natural\\ Language\end{tabular}}} &
\multicolumn{1}{c}{\textbf{\begin{tabular}[c]{@{}c@{}}Spatial\\ Position\end{tabular}}} &
\multicolumn{1}{c}{\textbf{\begin{tabular}[c]{@{}c@{}}Visual\\ Information\end{tabular}}} &
\multicolumn{1}{c}{\textbf{\begin{tabular}[c]{@{}c@{}}Monaural\\ Audio\end{tabular}}} \\
\midrule
\bf Primary Info & Semantic, Relational, Implicit Spatial & Explicit Spatial, Dynamic & Semantic, Spatial, Dynamic & Acoustic \\
\midrule
\bf Control Precision & Low & Very High &High & N/A \\
\bf Abstraction Level & High & Low & High & Low \\
\bf Interpretability & Indirect & Direct & Indirect & Indirect \\
\bottomrule
\end{tabularx}
\end{table}

\paragraph{Abstract Intent versus Geometric Precision}
A fundamental trade-off exists among the different input representations: the opposition between the level of abstraction in control and its precision. 
Natural language and visual information represent the pinnacle of abstract, human-centric control. 
Natural language provides an intuitive way to specify semantic content (e.g., "a bird is chirping") and relational spatial attributes ("on the left"). 
Similarly, visual information from images or videos offers rich spatial and semantic context. 
These inputs describe what exists in a scene and how its components are related, which aligns closely with human perception.

However, this intuitiveness comes at the cost of reduced precision. 
The system must infer precise physical parameters from abstract descriptions. 
The BAT model \cite{zheng2024bat} exemplifies this challenge, utilizing a large language model to interpret complex natural language queries regarding "sound event detection, direction and distance estimation, and spatial reasoning". 
This highlights a critical point: high-level abstract inputs require a sophisticated, AI-based interpretation layer to translate human intent into machine-executable instructions.

In contrast, spatial position data provides the highest degree of precision. Cartesian or spherical coordinates offer direct and unambiguous guidance for placing sound sources. 
This makes it indispensable for tasks requiring absolute accuracy, such as providing ground truth for training and evaluating sound localization models, or simulating precise physical phenomena like the Doppler effect by incorporating velocity vectors. 
The inherent trade-off is that this representation lacks semantic context and is non-intuitive and tedious for manually specifying complex acoustic scenes.

\paragraph{Monaural Audio as the Acoustic Core}
Unlike other inputs that primarily define where a sound is, monaural audio defines what the sound itself is. 
It constitutes the "foundational acoustic content" for many spatial audio tasks, providing core acoustic characteristics such as timbre of a specific instrument or the phonetic features of speech. 
Therefore, monaural audio plays a unique role in the ecosystem of input representations.

Many advanced generative systems follow a two-stage principle: first, a source model (such as AudioGen \cite{kreuk2022audiogen}  or AudioLDM \cite{liu2023audioldm}) generates a monaural audio stream; then, this stream is spatialized or upmixed into a multichannel or binaural format under the guidance of other input modalities, such as visual or positional data. 
This architecture clearly separates the problem of content generation from that of spatial rendering, enabling modular and flexible system design. 
Consequently, monaural audio is not an alternative option parallel to other input forms, but rather the fundamental substrate upon which they act.

\paragraph{Synergy and Multimodal Control}
The most powerful spatial audio systems are increasingly moving towards multimodality, creating comprehensive control schemes by combining the strengths of different input types to overcome the limitations of any single modality. 
The synergy between vision and audio is particularly potent. 
In audio-visual source separation tasks, the visual presence of an object (e.g., a speaking person) provides a strong, albeit implicit, cue for isolating its corresponding sound from a noisy mixture. 
In generation tasks, visual information can guide the spatialization process; for example, a U-Net architecture can take a monaural input and, guided by a video, render a spatially correct binaural or stereo output. 
The audio-visual matching task  is considered crucial, highlighting the deeply learned correspondences between these modalities.

Similarly, adding explicit spatial position data (such as source orientation and velocity) to a monaural audio stream allows for the simulation of highly realistic dynamic effects, like the Doppler shift , elevating realism to a level unattainable with static spatialization.

\subsubsection{Output Representations Comparison}
In the preceding sections, we have discussed the three spatial audio representation paradigms: channel-based, scene-based, and object-based. They represent an evolutionary path for spatial audio technology, moving from concrete to abstract and from rigid to flexible representations. To more intuitively illustrate their differences, connections, and respective trade-offs, this section provides a comprehensive comparative analysis of these three paradigms.

\begin{table}[ht]
\centering
\caption{Comparative Analysis of Spatial Audio Output Representations}
\label{tab: com_out}
\renewcommand{\arraystretch}{1.5}
\renewcommand{\tabularxcolumn}[1]{>{\centering\arraybackslash}m{#1}}

\begin{tabularx}{\textwidth}{
    >{\bfseries}l 
    X             
    X
    X
}
\toprule
\multicolumn{1}{c}{\textbf{Attribute}} & 
\multicolumn{1}{c}{\textbf{Channel-Based}} & 
\multicolumn{1}{c}{\textbf{Scene-Based}} & 
\multicolumn{1}{c}{\textbf{Object-Based}} \\
\midrule
Freedom of Listening Position & Limited & High & Moderate \\
Playback System Dependency & Very High & High & Low \\
Scalability & Low & Moderate & Excellent \\
Playback-End Complexity & Low & High & Moderate \\
\midrule
Common Formats & 
Stereo; 5.1, 7.1 Surround Sound & 
Ambisonics; Wave-Field Synthesis (WFS) & 
Dolby Atmos; DTS:X; MPEG-H 3D Audio \\
\bottomrule
\end{tabularx}
\end{table}

Table \ref{tab: com_out} presents a comparative analysis of the three primary spatial audio output representations. Each paradigm possesses unique advantages and limitations, making it suitable for different application scenarios and user requirements.

Playback system dependency and scalability are key to understanding the evolution of these three paradigms. 
Channel-based formats exhibit very high system dependency but poor scalability. 
This is because their audio mix is baked-in for a specific, standardized loudspeaker layout (e.g., 5.1 surround sound). 
Any playback system that deviates from this layout will degrade the intended spatial effect. 
In contrast, object-based formats feature low dependency and excellent scalability. 
They achieve this by decoupling the audio content from its metadata, which allows the playback device to render the audio in real-time according to its own arbitrary loudspeaker configuration. 
Consequently, a single master file can be adapted to any system. 
Scene-based formats occupy a middle ground. 
Their high dependency stems from the requirement for numerous loudspeakers and complex processing systems to physically reconstruct the sound field. 
Their moderate scalability is demonstrated by the ability to improve performance by increasing the system order (e.g., Higher-Order Ambisonics), though this significantly increases system cost and complexity. 

Freedom of listening position and playback-end complexity are directly related to user experience and implementation cost. 
Channel-based formats confine the listener to a narrow sweet spot, but their playback-end complexity is low, requiring only simple channel-to-loudspeaker mapping. 
Scene-based formats offer high freedom, allowing listeners to move freely within a designated area. 
However, this comes at the cost of very high playback-end complexity, which involves real-time decoding and substantial signal processing. 
Object-based formats provide moderate freedom of movement (depending on the rendering system). 
Their moderate to high playback-end complexity arises from the need for a real-time rendering engine to process metadata and dynamically generate the mix.

Overall, these three paradigms are not mutually exclusive; rather, each has its optimal application domain. 
Channel-based technology retains its place in traditional media due to its simplicity and broad compatibility. 
Scene-based techniques offer irreplaceable advantages in applications requiring high physical fidelity and large-scale public experiences. 
Meanwhile, object-based technology, with its unparalleled flexibility and interactivity, has become the core driver for next-generation immersive media, such as VR/AR, gaming, and streaming. 
Understanding their fundamental differences is crucial for selecting and implementing the most appropriate spatial audio solution.

\section{Understanding Approaches of Spatial Audio}
\label{sec: under}
\subsection{Overview}

Traditional audio understanding tasks encompass a range of sub-tasks such as speech recognition, speaker identification, and sound event detection, and are closely tied to joint learning with other modalities, including basic audio interpretation, text-to-audio understanding, and vision-to-audio understanding.

Spatial audio understanding extends traditional audio analysis by leveraging spatial cues to interpret complex acoustic scenes. A foundational task in this domain is \textbf{Sound Event Localization and Detection (SELD)}, which aims to simultaneously answer "what" sound is present and "where" it originates from. 
By utilizing the rich inter-channel information from multichannel or Ambisonic recordings, SELD systems perform joint Sound Event Detection (SED) and Direction-of-Arrival (DoA) estimation, forming the bedrock for most higher-level understanding applications.

Building upon the capabilities of SELD, \textbf{Spatial Audio Separation} emerges as a critical downstream task.
Its goal is to isolate individual sound sources from a complex mixture, effectively solving the "cocktail party problem." 
Unlike monaural separation, which relies solely on spectral patterns, spatial separation exploits binaural or multichannel cues (e.g., ITD and ILD) to distinguish and extract overlapping sources with much greater accuracy, making it essential for applications like hearing aids and immersive communication.

To achieve a more holistic and human-like perception, the field is increasingly moving towards \textbf{Joint Learning with multiple modalities}. 
This frontier of research integrates spatial audio with other streams of information. 
Key directions include aligning audio with visual information to understand scene geometry and object locations, modeling environmental acoustics to capture room characteristics like reverberation, and leveraging natural language with Large Language Models (LLMs) to reason about spatial relationships. 

Together, these research directions form a cohesive agenda for advancing spatial audio understanding tasks.

\subsection{SELD Tasks}

\subsubsection{Task Objective}
Sound Event Localization and Detection (SELD) systems support a broad array of machine-cognition tasks, including environment classification, sound-source localization, tracking of specific sound types, and acoustic event monitoring. 
The SELD task can be concluded as given an audio signal, the system outputs each source's class, spatial position and time interval.
In audio-visual task\cite{senocak2018learning}, SELD can likewise be framed as identify the sound-source class and its location within the frame. 
Over the years, research has progressed from classical signal/channel models and signal processing (SP) techniques to modern deep-learning-based methods.

SELD tasks relies on spatial audio, which preserves or recreates the three-dimensional character of natural sound via multichannel recordings or Ambisonics. 
Spatial-audio SELD comprises two subtasks: Sound Event Detection (SED), which determines the temporal intervals and classes of sound events, and Direction-of-Arrival (DoA) estimation, which predicts one or more source directions relative to a reference point (typically the microphone-array origin). 
Most studies focus on DoA—that is, the angular direction of a source—though some also address the source-to-array distance. 

By unifying temporal identification, spatial localization, and semantic labeling, SELD represents a paradigmatic audio-understanding task.

\subsubsection{Traditional Approaches}
Traditional sound source detection and localization methods rely primarily on signal processing and signal/channel models. 
These approaches typically employ time-frequency domain feature extraction. 

Common localization techniques include triangulation, trilateration, and multilateration. 
Parameters used to describe source locations include Time of Arrival (ToA), Time Difference of Arrival (TDoA), Frequency Difference of Arrival (FDoA), and Angle of Arrival (AoA), often referred to as Direction of Arrival (DoA). 
Corresponding methods encompass analysis of TDoA and FDoA which is frequently based on the Doppler effect and beamforming techniques that leverage the energy distribution across a microphone array. 
Signal processing algorithms applied within these frameworks include Minimum Variance Distortionless Response (MVDR), Delay-and-Sum (DAS), and Generalized Cross-Correlation (GCC).

\subsubsection{Deep Learning Approaches}
With the rapid advancement of deep learning, neural approaches have increasingly supplanted traditional signal-processing methods.
Works based on deep learning have been proposed to address the SELD task, including the use of convolutional neural networks (CNNs)\cite{Sun_2023_CVPR, hirvonen2015classification, cao2021improved}, convolutional recurrent neural networks (CRNNs)\cite{adavanne2018sound, nguyen2022salsa, shimada2022multi, yasuda2020sound, garcia2022binaural, nguyen2020sequence, shimada2021accdoa, krause2023binaural},
attention mechanisms\cite{tian2018audio, cao2021improved, santos2024w2v, kuang2022bast}, autoencoders\cite{huang2020time, wu2021sslide} and variational autoencoders\cite{bianco2020semi, bianco2021semi}.

Some works have considered the sound event detection and direction of arrival estimation as a joint task \cite{nguyen2022salsa, mesaros2019joint, cao2021improved,shimada2021accdoa}, 
while others separate the two tasks, \cite{Sun_2023_CVPR, pavlidi20153d, krause2023binaural, gan2019self, yasuda2020sound, may2010probabilistic, yang2022deepear} focus on SED task and 
\cite{tian2018audio, hirvonen2015classification} focus on DOA task.

A seminal binaural localization study by \cite{may2010probabilistic} used binaural recordings to estimate source azimuths. 
ITD were obtained by locating the peak of the cross-correlation function between left and right channel while ILD were incorporated to improve accuracy. 
A Gaussian mixture model (GMM) was employed to capture the joint dependency of ITD and ILD on azimuth. 
Although relatively simple by modern standards, this work delineated core requirements for effective sound-source localization—namely spectral analysis to extract spatial cues, and explicit handling of room effects, reverberation, and multi-source interference.

\cite{adavanne2018sound} introduced SELDnet, a classic deep-learning baseline that processes sound-event detection (SED) and direction-of-arrival (DOA) estimation in parallel using a convolutional recurrent neural network (CRNN). 
The SED branch outputs per-frame activity scores, with events deemed present if scores exceed a preset threshold—at which point the corresponding DOA is selected. 
The DOA branch formulates localization as a regression task, predicting the three-dimensional Cartesian coordinates of each active sound event.

Several works have focused on encoding representation and mapping methods to enhance performance.
\cite{pavlidi20153d} proposed a 3D DOA estimation method using sound intensity vector estimation. Signals from a spherical microphone array are transformed to the spherical harmonic domain; time-frequency regions with a single dominant source (SSZs) are identified, the intensity vector is estimated within each, and a smoothed 2D histogram of these estimates reveals the sources' DOAs.
\cite{rana2019towards} proposed a pipeline to predict the spatiotemporal location of sound sources in 3D space and time.
They introduced an automated Ambisonics estimation process based on deep-learning audio-visual feature embedding and prediction modules.
First, the system estimates the 3D location of sound sources, then uses this information to encode the audio in B-format.
They further proposed an Audio-Visual Distance Learning Network (AVDLN), which encodes data with pretrained CNNs, reduces audio and visual representations via two separate two-layer FC networks, and measures cross-modal distances using Euclidean distance.

The capture and reproduction of spatial audio have evolved to prioritize realism.
Beyond higher-order Ambisonics, binaural recordings now often use two microphones equipped with artificial pinnae to better mimic human hearing. 
This setup benefits applications such as hearing aids, which are constrained to two channels. 
\cite{yang2022deepear} propose DeepEar, a binaural-microphone localization system based on a multisector neural network. 
DeepEar divides space into discrete angular sectors, allowing simultaneous localization of multiple sources. 
First, a subtraction layer computes interaural feature differences. 
Then, three task-specific subnets process these features: SoundNet detects the presence of a source in each sector, AoANet regresses the source's azimuth angle, and DisNet estimates the distance from the ears to the source. 
The work also addresses monaural separation, acknowledging that some hearing-impaired users have only one functional ear. 
However, because it relies on fixed angular sectors, DeepEar's maximum number of concurrently localizable sources remains limited.

To address the challenge of an unknown maximum number of concurrent sources, \cite{kim2023ad} proposed an angular-distance-based multiple SELD model. 
Inspired by the YOLO multi-object detection framework, this approach does not require prior knowledge of how many sources may be active. Operating from an event-oriented perspective. 
It directly detects source positions on the spherical coordinate space, remaining robust to interference from multiple overlapping sound events.

The joint learning of SED and DOA tasks often leads to performance degradation in both subtasks.
\cite{hirvonen2015classification}jointly estimated spatial position and audio content type by classifying events into eight directional bins and two content categories (speech versus music). They demonstrated that a convolutional network combined with generic preprocessing can achieve strong results and be specialized for challenging conditions.
\cite{mesaros2019joint} augmented localization metrics with detection-related conditions and, conversely, incorporated location information into true-positive counts for detection. 
\cite{DBLP:journals/corr/abs-1905-00268} clarifies the relationship between SED and DOAE by showing that a network trained first for SED can reliably support DOAE. 
Their architecture features two branches: the SED branch is trained first, then its learned feature maps and true SED labels serve as a mask for the DOAE branch during training; at inference time, SED predictions provide the mask. 
This two-stage approach enhances each branch's representational power and preserves the contribution of SED to DOAE, avoiding the performance loss often seen in hard-parameter-sharing multi-task setups.
Another joint method, SALSA \cite{nguyen2022salsa}, is a joint SED and DOA feature for both FOA and MIC formats. SALSA (Spatial cue Augmented Log-Spectrogram) precisely maps signal energy to directional cues at the time-frequency level by stacking multichannel log-linear spectrograms with the normalized principal eigenvector of the spatial covariance matrix. Eigenvector normalization varies by array format to extract the appropriate spatial information.

\cite{cao2021improved} also build connections between SED and DOA tasks to address shortcomings in joint learning.
They further tackled the challenge of overlapping events of the same class by introducing a trackwise output format together with permutation-invariant training (PIT) at both frame and chunk levels, plus multi-head self-attention to separate tracks. 
They also adopted soft parameter-sharing in place of hard sharing, so joint SELD performance matches that of separately learned subtasks without compromise.
To address the challenge of balancing optimization objectives between sound event detection and DOA estimation, \cite{shimada2021accdoa} introduced the ACCDOA representation—an activity-coupled Cartesian DOA. In this scheme, each sound event's activity is encoded as the length of its corresponding Cartesian DOA vector. 
By unifying the SELD task into a single-target formulation, ACCDOA removes the need to trade off multiple losses and also reduces the overall network size.
Furthermore, their following work \cite{shimada2022multi} maintained the advantages of class-wise outputs by extending the ACCDOA representation to a multi-ACCDOA format and proposing auxiliary duplicating permutation-invariant training (ADPIT). 
By duplicating each ground-truth vector as an auxiliary target, each track in the multi-ACCDOA output learns more effectively.

Most DNN-based, data-driven DOA estimators excel on single sources but degrade sharply under overlap, whereas physics-based methods remain robust but less precise. 
\cite{yasuda2020sound} combined these paradigms by refining physics-based sound intensity vectors (IVs) with DNN-based denoising and source separation. 
Their system takes as input log-mel spectrograms, normalized Mel-domain IVs, and angular masks, and uses two networks—MaskNet to estimate T-F masks for denoising and source separation, and VectorNet to remove residual overlapping components such as reverberation—thereby achieving accurate DOA estimation even in overlapping scenarios.

\cite{krause2023binaural} present one of the few studies on listener motion in binaural localization.  
They conducted exhaustive experiments comparing a joint SDEL model with separate DOAE and SDE systems under multiple scenarios, demonstrating the benefits of motion-based cues.
While \cite{garcia2022binaural} investigates how head rotation affects DOA estimation in binaural sound source localization (SSL). 
They evaluate static-head baseline, network without rotation information, and variants that differ in how they process quaternions. 

Besides methods that jointly optimize SED and DOA, some works have decoupled the two tasks and then aligned their outputs with a modular, hierarchical design.
\cite{nguyen2020sequence} decouple the learning of sound event detection and direction-of-arrival estimation by adopting a two-stage strategy. 
The system performs event detection and DOA estimation independently in the first stage.
In the second stage, a deep neural network is trained to align the output sequences produced by the event detector and the DOA estimator. 

Existing methods often rely on contrastive learning, which can misclassify negative samples.
To enhance unsupervised audio-visual source localization,
\cite{Sun_2023_CVPR} introduced a self-supervised audio-visual source localization framework based on false-negative-aware contrastive learning. 
The main optimization objective is a noise-contrastive estimation (NCE) loss applied to audio-visual pairs.
\cite{santos2024w2v} also introduce a self-supervised model, w2v-SELD, which adapts the wav2vec 2.0 architecture to accept four-channel raw audio input. 
The model first generates feature vectors that are processed by Transformer blocks with multi-head attention. 
After the Transformer layers, it splits into two task-specific branches: one for sound event detection formulated as a multi-class classification problem, and one for direction-of-arrival estimation treated as a regression task.
It demonstrated the feasibility of building robust SELD models without relying on large-scale labeled spatial audio datasets.

Humans naturally combine audio and visual cues to localize objects. 
To overcome the limitations of manually labeled data in audio localization systems, \cite{gan2019self} leverage unlabeled audio-visual videos in a self-supervised framework, which transfers knowledge from an existing visual detection model into the audio domain.
They use a teacher network (YOLOv2) to predict multiple bounding boxes and class probabilities for moving vehicles in each frame. 
These detections serve as pseudo-labels to supervise a "student" audio subnet, StereoSoundNet, where audio segments are first converted into spectrograms, then encoded by convolutional layers and MLP, concatenated, and then decoded to restore spatial resolution. 
\cite{tian2018audio} introduced an Audio-Guided Visual Attention mechanism to capture audio-visual correlation and proposed the Dual Multimodal Residual Network (DMRN) to fuse modalities alongside an audio-visual distance learning network, which outperforms separate modeling.
DMRN updates audio and visual features jointly which preserving modality-specific information while injecting complementary cues and uses LSTMs for implicit temporal alignment. 

Recent advances have applied enhanced ResNet\cite{he2016deep}, Transformer \cite{vaswani2017attention}, and autoencoder (AE/VAE) \cite{rumelhart1986learning, kingma2013auto} architectures to audio event detection and sound-source localization, predominantly using semi-supervised or unsupervised learning to overcome the scarcity and high cost of labeled data. 
\cite{naranjo2020sound} demonstrated that integrating squeeze-and-excitation residual blocks into a CRNN improves SELD performance. 
To address CNNs' difficulty in capturing global acoustic context, \cite{kuang2022bast} proposed the end-to-end Binaural Audio Spectrum Transformer (BAST), which uses a dual-input hierarchical design with three Transformer encoders to mimic human auditory pathways. 
After separately processing the left and right channels, BAST fuses them via addition, subtraction, or concatenation and outputs source coordinates. 
Both parameter-sharing (BAST-SP) and non-sharing (BAST-NSP) variants significantly outperform CNN-based baselines.

AE and VAE based models have also been adapted to these tasks. 
\cite{huang2020time} introduced a time-domain, unsupervised localization method using autoencoders that removes the need for delay compensation or precisely aligned labels by learning acoustic transfer function filtering and inverse filtering mapping directly on raw multichannel signals. 
Their framework employs an "alternating and splitting" multi-task training strategy to enforce accurate source recovery and inter-channel consistency, and as a fully unsupervised end-to-end system, it's capable of generalizing to arbitrary microphone arrays.

\cite{wu2021sslide} presented SSLIDE, an AE architecture with one encoder and two parallel decoders. 
One for multipath alleviation and one for localization. 
During training, both decoders run, but at test time the multipath decoder is disabled. 
SSLIDE exhibits strong generalization across spatial locations, reverberation times, microphone perturbations, and recordings from unknown environments.

\cite{bianco2020semi, bianco2021semi} employ semi-supervised VAE approach for sound-source localization.
\cite{bianco2020semi} applied semi-supervised VAE learning to source localization by encoding the phase of the relative transfer function (RTF) between microphones into the latent space while jointly training a DOA classifier. 
Their VAE-SSL approach uses convolutional networks and variational inference to approximate the posterior distribution. 
Experiments show VAE-SSL outperforms CNN-based methods in label-limited scenarios and generalizes well to unseen source positions; 
similarly, \cite{bianco2021semi} further demonstrated the model's robustness to unseen source locations.
In addition, they showed that the generated RTF phase samples also reflect learned physical characteristics of the acoustic environment.

Many methods and model architectures have been applied to the SELD task in spatial audio, with the most commonly used structures being based on CNN, RNN, and hybrid models. 
More recent models, such as AE, VAE, and Transformer, have also been utilized for SELD tasks, as they are better at capturing time-frequency features from spectrograms. 
Overall, AI-based solutions have demonstrated superior performance in SELD tasks compared to traditional static algorithms and rule-based parameter methods. 
In particular, AI models excel in complex environments, where they can better adapt to intricate sound sources and reverberation, offering enhanced robustness.

\subsection{Spatial Audio Separation}
\subsubsection{Task Objective}
Audio source separation aims to recover the individual signals of underlying sound sources from a mixed audio recording. 
The two main categories of source separation are speech separation and music separation. 
Speech separation involves extracting a target speech signal from complex background noise or overlapping speakers, with applications in hearing-aid enhancement and noise-robust automatic speech recognition (ASR). 
Such problem is well known as cocktail party problem.
Music separation, by contrast, isolates singing voices and specific instruments from mixtures containing multiple accompaniments, enabling tasks such as vocal removal or instrument-specific extraction. 

Monaural separation methods have encompassed techniques based on mutual information \cite{fisher2000learning}, subspace analysis \cite{pu2017audio, smaragdis2003audio}, matrix factorization \cite{gao2018learning, parekh2017motion, sedighin2016two}, and correlated onsets \cite{barzelay2007harmony, li2017see}. 
Most existing systems integrate visual information to guide audio separation including speech separation \cite{afouras2018conversation, ephrat2018looking, owens2018audio}, musical instruments \cite{zhao2018sound}, and other objects \cite{gao2018learning}.  
Beyond purely audio-visual separation, recent work has also explored multimodal audio-separation frameworks that integrate additional modalities \cite{cheng2024omnisep}.

When audio is recorded with two or more than two channels, spatial cues between microphones can be exploited to improve separation.
However, such traditional methods degrade substantially when sources are closely spaced or heavily overlapped. 
Deep learning approaches often prioritize spectral information. 
Deep neural networks (DNNs) have become the dominant approach in audio source separation, typically modeling features in the Mel-spectrogram or short-time Fourier transform (STFT) domains. 
However, most work on multichannel separation remains limited in its ability to perceive three-dimensional visual scenes with spatial audio, and most approaches focus on audio separation with visual guidance.
For example, lip motion information \cite{gao2021visualvoice} and facial expression information \cite{lee2021looking} are used to separate the speech of different speakers.
Other methods \cite{chen2023iquery, tzinis2020into, tzinis2022audioscopev2} use multimodal attention to build the association between visual and audio modalities.

\subsubsection{Binaural Audio Separation}
Binaural audio separation aims to extract and reconstruct individual concurrent sound sources from the two-channel signal captured by a pair of microphones or a dummy head simulating human ears. 
Unlike monaural separation, it leverages not only spectral information but also spatial cues such as ITD and ILD to significantly improve the accuracy of separating and localizing overlapping sources. 
Traditional methods include deep clustering\cite{wang2018multi}, beamforming\cite{adel2012beamforming}, and independent component analysis (ICA). More recent approaches combine HRTFs or end-to-end deep neural networks, achieving substantial progress in challenging reverberant and multi-source scenarios. 

Binaural audio separation is widely applied in virtual reality, hearing aids, acoustic scene analysis, and interactive audio systems, delivering a more realistic and immersive listening experience.
Representative works on binaural-audio speech separation include \cite{zhang2017deep, weiss2009source, wang2018combining, han2020real, gao20192, ye2024lavss, deleforge2012cocktail}.

Fundamental machine learning techniques have long been applied to binaural audio separation tasks.
\cite{weiss2009source} combine a probabilistic model of the observed binaural level and phase differences with a prior statistical model of the sources, and derive an EM algorithm to estimate the maximum-likelihood parameters of the joint model. 
This system is capable of separating more sources than the number of observed channels.

In order to extend auditory processing toward human-robot communication, \cite{deleforge2012cocktail} introduced a novel method based on a generative probabilistic model and active binaural hearing, enabling a robot to robustly perform sound-source separation and localization. 
By exploiting interaural spectral cues within a constrained mixture model, their approach captures the richness of data gathered by two microphones mounted on a human-like artificial head, thus pioneering the unexplored field of sensorimotor learning for robot audition. 
This work integrates the robot's auditory and motor capabilities into a unified framework and demonstrates sound-source separation and localization in a realistic "cocktail-party" scenario, establishing a benchmark for sensorimotor learning in robot audition within the emerging field of human-robot interaction.

In reverberant environments, it remains challenging to separate multiple speakers using multichannel audio.
\cite{zhang2017deep} proposed and extracted a novel two-dimensional interaural time difference (ITD) feature to complement spectral cues, using the ideal ratio mask (IRM) as the training target, and demonstrated excellent separation performance in challenging multi-source and reverberant environments.
\cite{wang2018combining}  tightly integrate complementary spectral and spatial features within a deep learning framework to address this problem. 
Their method first employs a two-channel Chimera++ network to identify the time-frequency units dominated by the target speaker, performing direction estimation only on those units. 
This allows the enhancement network to be trained jointly on spatial and spectral features, thereby extracting the target speaker according to the estimated direction and specific spectral structures. 
By leveraging the phase enhancement provided by Chimera++-driven mask-based beamforming, the separation performance is further improved. 
However, the approach remains constrained by its assumption of a known speaker count in advance, and the use of bidirectional LSTMs makes it unsuitable for real-time online processing.

Several previous works focused on generating a single-channel output for each target speaker which discards the spatial cues needed for source localization. 
Preserving spatial information in speech separation is critical for accurate acoustic scene rendering.
\cite{han2020real}proposed a multi-input multi-output (MIMO) end-to-end extension of the time-domain audio separation network (TasNet) preserving the binaural cues of separated sources. 
It separates target speakers into two channels while achieving low latency and high fidelity and enables real-time acoustic scene modification and meets online processing requirements and also retains the perceived spatial cues.

2.5D Visual Sound \cite{gao20192} is a pioneering work in spatial audio generation: although its main goal is to convert monaural audio into binaural audio, it extracts spatial cues embedded in video and uses a multimodal approach to recover spatially correlated information. 
It focuses on the task of generating binaural audio while also demonstrating the model's ability to infer different sound-source positions from visual cues. 
It adopts a Mix-and-Separate strategy: the binaural tracks of two videos are mixed together, and the network is conditioned on each video's visual stream to separate out the corresponding binaural audio. 
Following the approach of \cite{zhao2018sound}, they use ratio mask and log-magnitude spectrogram. The model is trained using a per-pixel L1 loss.
Beyond binaural audio generation, experiments show that the self-supervised representations learned by this network can further improve audio-visual source separation.

To overcome the limitation of monaural audio-visual separation (MAVS) methods which focus solely on the acoustic characteristics of sound sources and lack any understanding of their spatial positions \cite{ye2024lavss} proposed Location-Guided Audio-Visual Spatial Audio (LAVSS) built upon the mainstream MAVS framework.
The proposed method first detects sounding objects to obtain region-level visual embeddings and then explicitly encodes each object's spatial position. 
Positional embeddings serve as additional guidance, revealing spatial cues that facilitate the separation of sources from different directions. 
Inter-microphone phase differences (IPD) are incorporated to characterize spatial features inherent in binaural audio.

The multi-scale attention-based fusion network integrates visual, positional, and audio features, allowing all modalities to work in concert and substantially improve audio-visual spatial audio separation.
This work introduced the concept of Audio-Visual Spatial Audio Separation (AVSS) for the first time and guiding the separation process through the fusion of object position representations with visual information.

\subsubsection{Multichannel Audio Separation}
Traditional audio separation in binaural scenarios often struggles when the number of sources exceeds the two available channels and when sound events overlap, since class-wise output formats become limiting in overlapping cases. 
Consequently, researchers have turned to multichannel audio separation and proposed various methods to address the underdetermined mixing problem.

Multichannel audio separation refers to techniques that exploit multi-channel signals—often captured with more complex array geometries or higher-order sound-field encodings such as Ambisonics—to decompose a mixture into its constituent sources. 
Unlike binaural separation, which relies solely on interaural time and level differences from two channels, multichannel approaches leverage richer spatial information and array geometry to improve separation accuracy through source localization, dereverberation, and adaptation to dynamic scenes.
However, multichannel separation can also suffer from spatial-filter breakdown and other challenges. In environments with overlapping sources and high reverberation, it must estimate a larger number of mixing parameters and source directions simultaneously. 
To overcome these difficulties, researchers have explored combinations of beamforming, blind source separation, and time-frequency masking techniques. At the same time, deep-learning-based methods have effectively driven further advances in efficient and precise multichannel audio separation.
Multichannel audio separation work has been explored in \cite{wang2018multi, morgado2018self,nugraha2016multichannel}.  

A classic example is multichannel spatial clustering via model-based source separation \cite{wang2018multi}, which, under a speech-sparsity assumption, uses Gaussian mixture models to cluster time-frequency units into sources based on interchannel time differences (ITDs), phase differences (IPDs), level differences (ILDs), and spatial spread, which are widely adopted in multichannel speech isolation and separation \cite{nakatani2021blind, yoshioka2018multi}.

\cite{nugraha2016multichannel} present a foundational DNN-based approach. 
They propose a framework for multichannel audio source separation that uses deep neural networks (DNNs) to model the source spectra and integrates a classical multichannel Gaussian model to exploit spatial information. 
Model parameters are estimated via an iterative EM algorithm and subsequently used to construct a multichannel Wiener filter.

Traditional spatial clustering methods, such as CACGMM, rely solely on phase or directional statistics, limiting their ability to capture the rich spectral structure of speech signals. 
Although early spectral-spatial integration approaches like DOLPHIN attempted to address this limitation by incorporating Gaussian Mixture Models (GMMs) for spectral modeling, GMMs are overly simplistic in capturing the complexity of speech spectra and often struggle to adapt under varying noise conditions.

In a more recent development, \cite{nik2024unsupervised} introduced an innovative approach that combines a Convolutional Variational Autoencoder (Conv-VAE) with an LSTM network to separately model local spatial features and global temporal dynamics. 
This architecture enables end-to-end unsupervised learning of Bayesian Surprise from spatial audio, facilitating robust detection of anomalies or attention-worthy events.

Similarly, \cite{zmolikova2021integration} proposed a unified framework that integrates time-frequency spectral information with spatial cues to recover clean speech signals from multiple speakers. 
Their method leverages Variational Bayesian inference to iteratively update the time-frequency masks, VAE latent variables, and spatial model parameters, achieving joint, end-to-end spatial audio separation.

\cite{wang2018multi} propose a method based on two channel audio separation which investigates the simultaneous integration of spectral and spatial features within a deep clustering framework to exploit complementary information for enhanced speech separation. 
A reference microphone is selected at run time and for each pair comprising the reference and a non-reference microphone, a two-channel deep clustering model extracts an embedding for every time-frequency unit. 
This procedure renders the algorithm directly applicable to microphone arrays with any number of elements and diverse spatial layouts.

Similar to 2.5D Visual Sound, \cite{morgado2018self} address the task of converting monaural audio into multichannel spatial audio while simultaneously separating individual sound sources and mapping them onto the spatial spherical surface.
In the final ambisonics generation step, the separated sources are weighted by localization weights to produce the ultimate spatial audio output. 
As with most audio separation approaches, their system employs a neural network guided by cross-modal video analysis, but it is not limited to conventional speech or music separation: it targets the extraction of multiple, unknown source types, and its separation module is trained without explicit supervision. 
Both the separation and localization modules are driven by a multi-modal audio-visual analysis module.
Because this method splits spatial audio generation into predicting attenuation maps for source separation and localization weights for source localization, it also accomplishes the audio separation task alongside the generation.

\subsection{Joint Learning}
\subsubsection{Task Objective}
Humans rely on multimodal cues when interpreting and engaging with real-world events \cite{jiang2024exploring} and the demand for high-quality audiovisual experiences has grown. 

In recent years, an increasing number of researchers have examined the relationship between audio and visual information and proposed various methods to leverage these modalities for joint learning. 
The visual modality has emerged as an important source of auxiliary information for audio processing, functioning both as a supportive cue and as a supervisory signal.

Also the acoustic environment contains rich spatial information, but this information is often implicit.
How to enable models to understand the unique spatial information represented by their environments is another important research direction.

Spatial information conveyed through natural language descriptions is also crucial: everyday spatial cues often appear in phrases, for example, "a car approaches from the left" inherently encodes a direction. 
Some researchers have begun to leverage these language-based spatial descriptions to aid audio processing or to employ large language models for interpreting such spatial cues.

\subsubsection{Alignment Between Audio and Visual Information}
Unlike monaural audio tasks, panoramic visual data and spatial audio both carry rich spatial cues, and aligning these two modalities is a key research challenge. 

\cite{morgado2020learning} introduces the audio-visual spatial alignment (AVSA) task, using contrastive learning to train a model to predict correct correspondences between audio and video. 
Their transformer-based architecture converts features between the audio and video modalities while integrating information from multiple viewpoints. They also adopt a curriculum learning strategy, first aligning at the full-video level and then refining alignment at the crop level. 
Experiments demonstrate that AVSA learns representations superior to prior self-supervised methods on downstream tasks such as audio-visual correspondence verification, video semantic segmentation, and action recognition.

\cite{yang2020telling} leverage the spatial correspondence between audio and video for self-supervised learning. 
Their work focuses on determining whether the left and right audio channels have been swapped in a video, thereby forcing the model to learn audio-video spatial alignment. 
The proposed spatial alignment network fuses visual and audio streams and then performs a binary classification. 
Its visual subnetwork, audio subnetwork, and fusion subnetwork each produce features that benefit downstream tasks. 
By extending this method to 360° video, the authors further demonstrate that the learned spatial audio representations effectively support sound localization and spatial understanding in panoramic video.

\subsubsection{Environment Information}
Beyond source information, the scenes and environments we inhabit also contain rich spatial acoustic cues, though these are often implicit. 
Faithful spatial audio reproduction typically requires knowledge of the acoustic environment. 
For instance, the sound field in a vast auditorium differs markedly from that in a small conference room with sound-absorbing materials.
In such varied acoustic settings, the propagation, reflection, and attenuation behaviors of sound likewise change. 
\cite{hameed2004psychoacoustic} investigates humans' ability to perceive virtual room size purely through auditory cues. 
Consequently, understanding these environmental acoustic characteristics is an important direction for spatial audio understanding research.

\cite{liang2023av} explore the synthesis of videos with novel viewpoints accompanied by spatial audio. 
They integrate audio-propagation priors obtained from an acoustic-perception module into Neural Radiance Field (NeRF) \cite{mildenhall2021nerf}. 
By tying the audio synthesis to the scene's geometric and material properties, their method models the sound field from the emitter's perspective and enables recorded audiovisual scenes to produce realistic, visually consistent spatial audio when rendered from new positions and angles.

\cite{luo2022learning} propose Neural Acoustic Fields (NAFs), which learn an implicit representation of audio propagation in indoor environments. 
NAFs operate in the time-frequency domain and condition the decoding of each impulse response on local geometric information at both the emitter and listener positions. 
This design allows the model to transfer local acoustic cues learned from training pairs to novel emitter-listener combinations.

Several studies focus on characterizing indoor room acoustics through simulated room impulse responses (RIRs). 
For instance, \cite{savioja2015overview, ratnarajah2024av, ratnarajah2022mesh2ir, bryan2020impulse, iso20093382, coldenhoff2024multi, majumder2022few, srivastava2021blind} examine room-acoustic parameters—such as geometry and surface materials—from a multichannel speech perspective. 
Others, like \cite{chen2024real}, investigate the generalization from simulated to real recordings, offering new strategies to overcome the scarcity of real-world spatial audio datasets.

\subsubsection{Visual Segmentation and Depth Estimation}
Visual inputs not only provide supervisory signals for object detection and source identification but also offer auxiliary information for audio depth estimation. 
In speaker recognition and instrument separation scenarios, the visual model detects and segments sound sources within the image and estimates their depth, thereby computing their precise 3D positions and supplying spatial cues essential for generating realistic spatial audio.

Motivated by extracting spatial information from vision cues, 
\cite{liu2025visual} also focuses on converting monaural audio into a binaural format to simulate 3D sound effects, with the goal of precisely synchronizing the audio with the spatial positions of visual elements.
They integrate YOLOv8 \cite{varghese2024yolov8} object detection with the Depth Anything \cite{yang2024depth} monocular depth estimation model to extract high-precision visual cues, thereby enhancing the accuracy of spatial audio alignment and enabling real-time automatic localization of audio sources in 3D space. 
The overall architecture comprises separate visual and audio processing pipelines rather than a jointly trained multimodal network.
In the visual pipeline, YOLOv8 detects objects and provides their 2D image coordinates, which are combined with depth estimates from Depth Anything to compute each target's full 3D position.
The audio pipeline begins with source separation: Conv-TasNet \cite{luo2019conv} is applied to speech separation tasks, while Demucs \cite{defossez2019demucs} is used for music separation.
Once individual audio streams are obtained, they are aligned to their corresponding visual detections. 
To balance between channels and strengthen spatial effects, the separated tracks undergo two spatialization methods: convolution with HRTFs and a 3D audio-positioning algorithm that simulates sound propagation for left-right, up-down, and front-back positioning.
Although functioning primarily as an intermediate spatial-audio synthesis module, this system leverages visual depth and scene information as effective supervision, underscoring a deeper understanding of the spatial characteristics inherent in audio signals.

\subsubsection{Natural Language Guided}
In addition to visual cues, humans rely on spatial information conveyed by natural language descriptions when interpreting spatial audio. 
Although these linguistic descriptions are not as direct as visual information, they nevertheless provide valuable spatial signals. 
Spatial information can be extracted from text embeddings and used as a cue to guide alignment and supervision in audio processing tasks.
Also, in recent years, the rapid advancement of large language models (LLMs) has prompted researchers to explore their use in audio processing tasks. 
LLMs can understand and generate natural language and, in multimodal settings, handle inputs such as images and audio.

Existing audio foundation models lack spatial awareness: they are typically trained on non-spatial audio-text pairs and therefore seldom learn how to perceive space.
In sound event localization and detection tasks, models cannot generate natural-language descriptions of sound positions, making it difficult to link spatial audio with text.
To address this, \cite{devnani2024learning} introduces ELSA. 
First, they propose a spatially augmented open-source audio dataset by simulating room acoustics and generating first-order Ambisonics recordings. 
They use a large language model to rephrase the original captions and also collect a small real-world spatial audio dataset for validation.
ELSA comprises an audio encoder paired with a text encoder that jointly learns both semantic and spatial attributes through contrastive learning. 
The audio encoder has two parallel branches—one for semantic content and one for spatial properties. 
During training, ELSA optimizes three additional spatial regression objectives (arrival direction in 3D space, source-receiver distance, and room size) alongside a batched contrastive loss to align audio and text embeddings.
Experiments show that ELSA successfully aligns spatial audio with its text descriptions and captures both semantic and spatial information effectively.
ELSA is considered as a representative work for alignment between spatial audio and text.

\cite{zheng2024bat} address the gap in using LLMs to infer sound direction and distance by proposing the BAT model. 
They first create the SPATIALSOUNDQA dataset, which contains natural-language question-answer pairs for sound event detection, direction and distance estimation, and spatial reasoning.
The QA pairs are then paraphrased with GPT-4 \cite{achiam2023gpt} to enhance diversity. 
BAT's architecture fuses a SPATIAL-AST audio-spectrogram transformer with LLaMA-2 7B \cite{touvron2023llama2}: a projection module maps SPATIAL-AST outputs into LLaMA-2's embedding space, and then fine-tuned in stages on different QA types to incrementally improve performance. 
Experiments show that BAT performs well on event detection and distance prediction, with the best results achieved using binaural inputs augmented by interaural phase differences (IPD) and a two-stage training regime. 
BAT is a pioneering work demonstrating LLMs' potential for spatial audio reasoning and opens a new direction for natural-language-driven spatial audio tasks.
\section{Generation Approaches of Spatial Audio}
\label{sec: gener}
As early as 1881, Ader conducted the first demonstration of spatial (stereo) sound by transmitting audio signals through a pair of microphones and telephone receivers. Early approaches to spatial audio generation primarily relied on conventional digital signal processing techniques, wherein each sound source was recorded individually and later mixed using various audio processing techniques in conjunction with predefined scene configurations and spatial parameters. However, such methods suffer from inherent limitations, including their reliance on extensive audio data and environmental information, as well as their poor adaptability to dynamic scenes and varying sound sources. Consequently, spatializing monaural audio has remained a prominent research challenge.

In recent years, the rapid advancement of generative models has significantly propelled the development of spatial audio generation techniques. End-to-end spatial audio generation methods, in particular, have continuously evolved, enabling the production of high-quality spatialized audio. As a relatively specialized subfield, the research on spatial music generation has preceded that of audio generation, achieving stereo music synthesis under more supervised conditions and with additional input like genre or style.

This section provides an overview of the recent progress in spatial audio generation, focusing on  both classical digital signal processing (DSP) techniques and recent deep learning-based architectures, and organizes the section according to model architecture.

\subsection{Traditional Spatial Audio Generation Methods}
Traditional spatial audio methods primarily focus on binaural audio, aiming to replicate the spatial perception of human hearing. 
Monophonic-to-binaural conversion seeks to recover the interaural time-difference (ITD) and interaural level-difference (ILD) cues that underpin spatial hearing. Early approaches relied on digital signal processing rules (DSP) \cite{brown1998structural, zhang2017surround, jianjun2015natural, moller1992fundamentals} or measured head-related transfer functions (HRTFs)\cite{guezenoc2020hrtf}to impose fixed delays and filters on single-channel inputs. 
Although these methods are intuitive and mathematically straightforward, they typically lack perceptual authenticity in dynamic scenes and fail to generate audio that closely resembles recorded binaural audio.
At the same time, with the increasing popularity of microphone arrays and ambisonic microphones, spatialization methods based on multichannel audio have also emerged. However, these traditional methods often struggle to capture the relationships between multiple channels and are gradually being supplanted by deep learning-based approaches.

\subsection{Deep Learning Based Spatial Audio Generation}
In recent years, rapid advances in deep learning have driven significant progress in spatial audio generation. In particular, neural network-based methods have achieved remarkable results in monaural-to-binaural audio conversion. 
Furthermore, the introduction of generative frameworks, like U-Net, diffusion models and variational autoencoders (VAEs), have opened new avenues for end-to-end synthesis of multichannel and ambisonic audio.
A summary of recent deep learning models is presented in the Table \ref{tab: gen}.

\begin{table}[ht]
    \centering
    \small
    \caption{Comparison of current spatial audio generative models. FOA means first-order ambisonics, while Multi denotes multi-channel audio. The input/output format in parentheses indicates optional ones. CNN based models are not included in this table.}
    \label{tab: gen}
    \scalebox{1}{
    \begin{tabular}{lcccccc}
        \toprule
        \textbf{Model} & \textbf{Input Format} & \textbf{Output Format} & \textbf{Architecture} \\
        \midrule
        \citealt{xu2021visually} & Mono, Image & Binaural & \multirow{8}{*}{Diffusion} \\
        Binauralgrad \cite{leng2022binauralgrad} & Mono & Binaural & \\
        Mo$\hat{u}$sai \cite{schneider2023mo} & Text & Binaural &  \\
        See-2-Sound \cite{dagli2024see} & Image, (Text) & Multi &  \\
        \citealt{evans2024long} & Text, (Audio, Duration) & Binaural &  \\
        DualSpec \cite{zhao2025dualspec} & Text & Binaural & \\
        ImmerseDiffusion \cite{heydari2025immersediffusion} & Text, (Position) & FOA & \\
        SonicMotion \cite{templin2025sonicmotion} & Text, Position & FOA & \\
        \midrule
        \citealt{kim2019immersive} & 360$^\circ$ Image & IRs & \multirow{3}{*}{Latent Diffusion} \\
        \citealt{huang2022end} & Mono, Position & Binaural & \\
        \citealt{yang2022upmixing} & Binaural/Multi & Multi & \\
        \midrule
        \citealt{lee2023neural} & Mono, Position, Orientation & Binaural & \multirow{5}{*}{Transformer} \\
        MusicGen \cite{copet2023simple} & Text & Mono/Binaural &  \\
        Ambisonizer \cite{zang2024ambisonizer} & Mono/Binaural & FOA &  \\
        ViSAGe \cite{kim2025visage} & Video, Camera Position & FOA &  \\
        ISDrama \cite{zhang2025isdrama} & Video, Audio, Text, Position & Binaural &  \\
        \midrule
        OmniAudio \cite{liu2025omniaudio} & 360$^\circ$ Video & FOA & \multirow{2}{*}{Flow Matching} \\
        Diff-SAGe \cite{kushwaha2025diff} & Class Label, Position & FOA &  \\
        \midrule
        Listen2scene \cite{ratnarajah2024listen2scene} & 3D Mesh, (Source \& Listener Position) & Binaural IRs  & \multirow{2}{*}{GANs} \\
        SAGM \cite{li2024cross} & Mono, Video & Binaural & \\
        \bottomrule
\end{tabular}}
\end{table}
\subsubsection{CNN Models}
Convolutional Neural Networks (CNNs) \cite{lecun1998gradient} are foundational architectures in deep learning, widely adopted for audio generation tasks due to their success in image processing.
While in spatial audio generation, CNNs are primarily employed to extract features from monophonic audio or visual information, leveraging convolutional operations for audio synthesis.

U-Net \cite{ronneberger2015u} was originally proposed for image segmentation and consists of a symmetric encoder-decoder structure. It inherits the core components of CNNs and introduces skip connections to efficiently extract and fuse multi-scale features, making it widely applicable to audio, temporal data, and multimodal tasks.
In the domain of visually guided binaural audio generation, several studies including \cite{gao20192, zhou2020sep, zheng2022interpretable, liu2024visually, garg2021geometry, li2021binaural, rachavarapu2021localize, zhang2021multi} adopt this U-Net backbone to map monophonic audio spectrograms and video frames to stereo outputs. 
Typically, a ResNet or similar convolutional network first extracts visual features from each video frame; these features are then either flattened or re-weighted via an attention mechanism and injected into the U-Net encoder or bottleneck, where they are fused in parallel with the audio representations.
The modular design of U-Net naturally accommodates the "conditional injection" of visual information, allowing for parallel or serial fusion of visual features at different levels to help the network learn precise spatiotemporal correspondences.

Although these methods share the same architectural foundation, they address distinct challenges such as limited annotated datasets, poor interpretability, and cross-modal misalignment, through a variety of innovations
\cite{zheng2022interpretable} introduces an interpretable learning target to overcome the lack of explainability inherent in traditional Difference Mask approaches.
To achieve more effective audio-visual alignment, 2.5 D Visual-Sound \cite{gao20192} propose MONO2BINAURAL, the first U-Net-based model that analyzes a monophonic audio track alongside its corresponding video frame to predict left and right channels for binaural reproduction.
Sep-Stereo \cite{zhou2020sep} devises an Attentive Pyramid Network (APNet) that fuses multi-level visual and audio features via convolutional layers, enabling the model to learn tight correspondences between object positions in the video and spectral content in the audio—thereby leveraging large-scale monophonic datasets for stereo synthesis.
To ensure that the synthesized audio exhibits both spatial continuity and perceptual realism, \cite{garg2021geometry} incorporate room geometry priors by adding an RIR-generation decoder and a spatial-consistency encoder external to the U-Net.
\cite{li2021binaural} introduces a cross-modal co-attention mechanism predicated on spatial consistency (i.e., left-right correspondence), facilitating semi-supervised and self-supervised training that reduces reliance on annotated binaural data. 
To provide stronger supervision, this work also defines an auxiliary task of determining whether an audio sample has been channel-swapped, sharing visual features between tasks but using separate attention modules to extract task-specific representations.
\cite{liu2024visually} extend the U-Net backbone with two lateral audio decoders, each conditioned on different visual features, and employ a self-supervised learning scheme to alleviate the need for extensive real binaural video data during training.
L2BNET \cite{rachavarapu2021localize} generates interaural difference signals by fusing audio with positional visual cues through an attention-based fusion module; it further pioneers the use of a “sound localization accuracy” criterion as an indirect supervisory signal, offering a novel multi-task framework that can also be employed to synthesize binaural datasets.
MAFNet \cite{zhang2021multi} introduces a multi-attention fusion network grounded in self-attention, designed to extract source-related spatial features from video frames and effectively integrate them with audio features.
Unlike the aforementioned works, \cite{lim2024enhancing} focus on the challenge of directly generating FOA-format audio from multiple modalities. By introducing a Channel Panning loss, they explicitly enhance the consistency of interaural and vertical sound energy distributions, thereby improving localization performance and demonstrating the efficacy of U-Net in high-level spatial audio generation tasks.

Collectively, these works demonstrate the remarkable flexibility of U-Net for spatial audio generation and, through diverse architectural enhancements and auxiliary tasks, effectively mitigate issues of dataset scarcity and cross-modal inconsistency.

360° visual inputs typically require specialized preprocessing before being used for audio-visual fusion in spatial audio generation, and convolutional neural networks (CNNs) are the prevailing choice for handling such panoramic data. 

Both \cite{kim2019immersive, rana2019towards} focus on extracting spatial cues from 360° video using CNN-based deep learning approaches. 
Specifically, \cite{kim2019immersive} model room geometry directly from panoramic imagery by leveraging depth estimation and semantic annotation to faithfully reconstruct real-world architectural features and synthesize corresponding room impulse responses (IRs). 
In contrast, \cite{rana2019towards} support both equirectangular and cubemap projections: they first extract visual embeddings with a pretrained VGG-19 and audio embeddings with a VGG-like network; these features are then processed by an SsM self-supervised module and integrated via an attention mechanism to produce a 3D localization probability map, which is ultimately used—via spherical-coordinate weighting—to generate FOA audio.
\cite{morgado2018self} also utilize a U-Net architecture to synthesize FOA audio from 360° video, incorporating self-supervised learning to enhance the model's performance.

Sonic4D \cite{xie2025sonic4d} adopts a modular architecture that first extracts semantically aligned dynamic visual scenes and raw audio from monocular video, providing foundational visual and acoustic priors for subsequent spatial audio synthesis.
In the second stage, it estimates the 3D trajectory of sound sources within a 4D scene (i.e., their 3D coordinates at different timestamps) through a process that includes 2D source localization, 3D coordinate back-projection, and trajectory optimization, thereby furnishing accurate source positioning for physics-based spatial audio simulation. 
Finally, it employs the gpuRIR tool based on the Image Source Method (ISM) to convolve monophonic audio with room impulse responses (RIRs), generating binaural audio outputs.
This phased approach effectively addresses the challenges of visual-audio-spatial alignment, culminating in immersive spatial audio generation for 4D scenes.

AV-Cloud \cite{chen2024av} addresses the challenge of real-time rendering of high-quality spatial audio for 3D scenes, ensuring seamless synchronization with visual content.
Its core innovation lies in constructing an audio generation pipeline that operates in parallel and independently of visual rendering.
It begins with sparse Structure from Motion (SfM) point clouds extracted from video, defining a set of audio-visual anchor points as a unified representation of the scene. Each anchor point is associated with 3D coordinates, RGB visual features, and audio effect embeddings.
The paper introduces an innovative Audio-Visual Cloud Splashing (AVCS) module that dynamically decodes contributions from these anchor points based on the listener's viewpoint and "splashes" them into a spatial audio transfer function.
Subsequently, the Spatial Audio Rendering Head (SARH) utilizes this function to transform a monophonic reference audio source into stereo audio that matches the viewpoint.
Since the entire process does not require waiting for image rendering, AV-Cloud can efficiently generate high-quality spatial audio that is synchronized with visual content.

These works collectively demonstrate the powerful capabilities of CNNs in spatial audio generation, particularly in effectively handling the fusion of visual information and audio features.

\subsubsection{Diffusion-based Models}
In recent years, diffusion models have emerged as a powerful generative paradigm, achieving state-of-the-art results in image and text generation, and their application to audio synthesis, particularly spatial audio, has grown rapidly.

Leveraging the Latent Diffusion Model framework, SpatialSonic \cite{sun2024both} pioneers end-to-end spatial audio generation by conditioning on both textual and visual inputs to extract spatial cues. 
Similarly, ImmerseDiffusion \cite{heydari2025immersediffusion} employs an end-to-end diffusion pipeline that integrates textual descriptions and spatial parameters, producing four-channel FOA format audio via a dedicated decoder.
SEE-2-Sound \cite{dagli2024see} introduces a zero-shot, diffusion-based approach: it first applies existing segmentation and depth-estimation networks to extract visual context, then uses the novel CoDi generator to synthesize monaural audio, and finally allocates channels to assemble a spatial audio output—its modular design confers strong extensibility. 
In the realm of music generation, Mo$\hat{u}$sai \cite{schneider2023mo} employs a two-stage cascading diffusion technique, combining a diffusion autoencoder for efficient compression with a cross-attention mechanism for text alignment, thereby achieving high inference speed and long-duration synthesis. 
Furthermore, \cite{evans2024long} propose a Diffusion-Transformer (DiT) based model for long-duration music generation, enhancing the temporal modeling capabilities of diffusion processes through a Transformer architecture while maintaining high audio quality, enabling coherent generation of extended musical pieces or humming.
DualSpec \cite{zhao2025dualspec} utilizes a VAE to obtain latent representations, which are then diffused in the latent space before being decoded into Mel and STFT spectrograms using the VAE decoder.
To address the limitations of existing FOA generation methods like Diff-SAGe \cite{kushwaha2025diff} and ImmerseDiffusion \cite{heydari2025immersediffusion} that only support static sound sources or stereo audio, SonicMotion \cite{templin2025sonicmotion} achieves end-to-end generation of dynamic FOA audio, enabling control via both text and spatial parameters for the first time.

These works collectively highlight the potential of diffusion models in spatial audio generation, particularly in handling multimodal inputs and end-to-end synthesis.

\subsubsection{Transformer-based Models}
Transformer \cite{vaswani2017attention}, renowned for its ability to model long-range dependencies, has become a popular backbone for generative models, initially revolutionizing natural language processing. 
Its capacity to capture global context has naturally extended its adoption to spatial audio generation.

Traditional pipelines treat "monaural-to-stereo" and "stereo-to-surround" as separate tasks. 
To address the scarcity of FOA-format audio, Ambisonizer \cite{zang2024ambisonizer} introduces a Transformer-based encoder-decoder architecture that, in conjunction with a VAE for extracting spatial location features, unifies monaural and stereo upmixing into a single FOA generation framework. 
While ViSAGe \cite{kim2025visage} integrates CLIP-derived visual embeddings, a Transformer-based audio codec, and camera-angle augmentation to perform end-to-end FOA synthesis, yielding high-fidelity audio with robust spatial consistency.
Further expanding multimodal capabilities, ISDrama \cite{zhang2025isdrama} accepts heterogeneous inputs including video, audio, text, and spatial position to employ a Transformer-driven generator to produce binaural Mel spectrograms that jointly model emotional prosody and spatial positioning. 
It leverages the Flow-Mamba architecture, which combines Transformer attention with Mamba's efficient long-sequence processing, enabling rapid and stable generation of expressive, spatially accurate speech.

Beyond speech, efficient generation of high-quality spatialized music is a growing research frontier. 
MUSICGEN \cite{copet2023simple} employs a single-stage Transformer with an interleaved token scheme to synthesize music with minimal computational overhead; its stereo output is achieved by reusing a simple interaural delay pattern on the monaural waveform, thereby generating binaural audio at no additional cost.

\subsubsection{Latent Diffusion Models}
Autoencoders (AEs) \cite{hinton2006reducing} and variational autoencoders (VAEs) \cite{kingma2013auto} are unsupervised learning frameworks that obtain low-dimensional latent representations by learning a compression-reconstruction mapping. 
Their architectural simplicity, end-to-end trainability, and strong capacity to model complex distributions have made them widely adopted for feature extraction in audio generation tasks.

Building on the SoundStream encoder, \cite{huang2022end} introduce a binaural decoder conditioned on source position and orientation, further enhanced through adversarial training to improve synthesis quality. 
\cite{yang2022upmixing} extend the VAE paradigm by disentangling spatial location and content information within music, enabling flexible upmixing from stereo to multichannel formats (5.1 surround). 
Their approach permits seamless transfer of instrument spatial distributions between songs and the automated generation of novel instrument placements, facilitating rapid, creative surround-sound production.

DualSpec \cite{zhao2025dualspec} combines a VAE that compresses both Mel-spectrograms and STFT representations into a shared latent space with a subsequent diffusion model to jointly optimize audio fidelity and spatial precision. 
As a hybrid architecture, OmniAudio \cite{liu2025omniaudio} refines the StableAudio framework \cite{evans2024fast} by initializing weights and adapting transform operations to effectively encode the four FOA channels thus providing robust feature representations for downstream spatial audio synthesis.

\subsubsection{Flow-based Models}
Flow matching \cite{lipman2022flow} is an emerging generative modeling paradigm that learns a time-dependent vector field transporting a simple base distribution to the data distribution via supervised regression on interpolation trajectories. 
By obviating the need for costly simulation during training, flow matching enables fully parallel, end-to-end optimization, thereby markedly accelerating model training on large-scale datasets which  makes it well-suited for a wide range of generative tasks.

Kushwaha et al. \cite{kushwaha2025diff} adopt the flow-matching framework to synthesize spectrogram representations directly from noise, resulting in an end-to-end pipeline for FOA audio generation. 
Similarly, OmniAudio \cite{liu2025omniaudio} integrates a flow-matching based generator, through self-supervised pretraining on extensive non-spatial audio data and a small FOA corpus, it learns universal audio patterns and achieves the first fully end-to-end conversion from 360° video to FOA-format audio.

\subsubsection{GANs based Models}
Generative adversarial networks (GANs) leverage an adversarial training framework to learn the true data distribution and produce high-fidelity samples, while flexibly incorporating visual, geometric, or other spatial cues. 

Listen2Scene \cite{ratnarajah2024listen2scene} adopts a hybrid GNN-cGAN architecture: a graph neural network encodes scene structure to extract spatial information, and a conditional GAN synthesizes audio, where the discriminator's ability to distinguish real from generated sound compels the generator to improve realism.
SAGM \cite{li2024cross} guides monaural-to-binaural audio generation using visual features extracted from video frames. By fusing AudioNet and VideoNet embeddings within a GAN framework and alternately updating generator and discriminator with enhanced visual context, SAGM achieves bidirectional, complementary learning and overcomes the limited visual guidance of prior methods.

\subsubsection{Other Models}
AV-NeRF \cite{liang2023av} employs a multi-layer perceptron (MLP) architecture that integrates both audio and visual information. Given a video recording of an audiovisual scene, it synthesizes new videos with spatial audio along any arbitrary camera trajectory within that scene, effectively aligning the listener's audio perspective with the visual viewpoint.
\section{Dataset and Evaluation of Spatial Audio}
\label{sec: data}

\subsection{Datasets}
Spatial audio data exists in a variety of formats, each reflecting different characteristics and tailored to specific tasks. This section provides an in-depth analysis of existing spatial audio datasets, illustrating the diverse methods of data collection and processing, and explaining how these elements contribute to the understanding of spatial audio.

Due to variations in recording equipment and application scenarios, spatial audio data comes in multiple formats, often accompanied by annotations and auxiliary data from other modalities. Moreover, because recording spatial audio is typically costly and resource-intensive, many existing approaches resort to using simulation systems to generate synthetic data from current monaural audio datasets. Some datasets also include real-world spatial audio crawled from the YouTube platform. The following subsections focus on the acquisition and processing methods—both recorded and simulated—associated with various spatial audio formats, including multi-channel audio, First-Order Ambisonics, and binaural audio. A summary of commonly used datasets is presented in the Table \ref{tab: data}.

\begin{table}[ht]
\centering
\small
\caption{Comparison of current spatial audio datasets. FOA means first-order ambisonics, while Multi denotes multi-channel audio.
}
\label{tab: data}
\scalebox{1}{
\begin{tabular}{lcccccc}
\toprule
\bf Dataset & \bf Format & \bf Collect & \bf Hours & \bf Type & \bf Labels \\
\midrule
Sweet-Home \cite{vacher2014sweet}  & Multi& Recorded & 47.3& Speech & Text \\
Voice-Home \cite{bertin2016french}  & Multi& Recorded & 2.5& Speech &  Text, Geomrtric \\
YT-ALL \cite{morgado2018self}& FOA & Crawled & 113 & Audio & Video, Text \\
REC-STEEET \cite{morgado2018self} & FOA & Recorded & 3.5 & Audio & Video \\
FAIR-Play \cite{gao20192} & Binaural & Recorded & 5.2 & Audio & Video \\
SECL-UMons \cite{brousmiche2020secl} & Multi & Recorded & 5 & Audio & Text, Geometric\\
YT-360 \cite{morgado2020learning} & FOA & Crawled & 246 & Audio & Video \\
EasyCom \cite{donley2021easycom} & Binaural & Recorded & 5 & Speech & Geometric, Text\\
Binaural\cite{richard2021neural} & Binaural & Recorded  & 2& Speech & Geometric \\
SimBinaural \cite{garg2023visually} & Binaural & Simulated & 116 & Audio & Video, Geometric \\
YouTube-Binaural \cite{garg2023visually} & Binaural & Crawled & 27 & Audio & Video\\
Spatial LibriSpeech \cite{sarabia2023spatial} & FOA & Simulated & 650 & Speech & Text, Geometric \\
STARSS23 \cite{shimada2023starss23} & FOA & Recorded & 7.5 & Audio & Video, Geometric \\
YT-Ambigen \cite{kim2025visage} & FOA & Crawled & 142 & Audio & Video\\
BEWO-1M \cite{sun2024both} & Binaural & Simulated & 2.8k & Audio & Text/Image, Geometric\\
MRSDrama \cite{zhang2025isdrama} & Binaural & Recorded & 98 & Speech & Text, Video, Geometric\\
\bottomrule
\end{tabular}}
\end{table}

\subsubsection{Multi-Channel Audio Datasets}

The most straightforward method for capturing spatial audio involves the use of multiple microphones, enabling the creation of immersive acoustic scenes that convey a sense of spatiality without requiring specialized recording hardware. High-quality, realistic multi-microphone corpora play a crucial role in narrowing the performance gap between close-talk conditions and far-field speech interaction.

Several corpora have been developed to support speech enhancement and automatic speech recognition (ASR) in reverberant or domestic settings. The REVERB Challenge corpus \cite{kinoshita2016summary} contains utterances from stationary speakers in reverberant rooms, recorded with a multi-channel circular microphone array, and is primarily intended for advancing speech enhancement techniques. The DIRHA corpus \cite{ravanelli2015dirha} includes both real and simulated multi-channel recordings across various home environments such as living rooms and kitchens. Simulated data are generated through a contamination approach \cite{matassoni2002hidden,ravanelli2012impulse} that combines clean speech, real background noise, and impulse responses (IRs) measured at corresponding locations. Similarly, the Sweet-Home corpus \cite{vacher2014sweet} offers multi-modal recordings, including distant speech captured during natural domestic activities via distributed microphone arrays, supporting research in ambient intelligence and smart home applications.

To facilitate research in spatial hearing and sound source localization, several datasets provide multi-channel recordings with accurate spatial annotations. The Voice-Home corpus \cite{bertin2016french} includes annotated speech recordings with transcriptions, along with corresponding room impulse responses under various acoustic conditions, enabling detailed investigations into spatial perception in domestic environments. The SECL-UMons dataset \cite{brousmiche2020secl}, recorded across multiple positions in office spaces, is designed for sound event classification and localization. To address the scarcity of labeled data for speaker tracking in smart home scenarios, the AVRI dataset \cite{qian2022audio} provides synchronized multichannel audio and video data with precise 3D positional annotations in reverberant indoor environments, supporting deep learning-based speaker localization.

More recent efforts have focused on dynamic or complex acoustic environments. The Wearable SELD dataset \cite{nagatomo2022wearable} features 24-microphone recordings mounted on a head-and-torso simulator (HATS), simulating pedestrian scenarios for sound event localization and detection in mobile contexts. The RealMAN dataset \cite{yang2024realman} further expands the research landscape by offering high-channel-count recordings with detailed transcriptions and geometric annotations. It includes speech playback from loudspeakers under diverse conditions—including indoor, outdoor, semi-outdoor, and transportation environments—mixed with various background noises. This corpus is designed to support robust modeling for both speech enhancement and multi-microphone source localization.

\subsubsection{First-Order Ambisonics Datasets}

First-Order Ambisonics (FOA) datasets have become a foundational resource for spatial audio research, particularly in tasks such as 3D sound localization, acoustic scene analysis, immersive audio rendering, and spatial audio generation. FOA captures a full-sphere representation of the sound field using four channels (W, X, Y, and Z) corresponding to the omnidirectional and directional components of the sound pressure. This format offers a compact yet spatially informative encoding of auditory scenes, making it especially suitable for machine learning applications in spatial hearing and audio-visual scene understanding. 

Due to the high cost of recording FOA data, scraping FOA audio and corresponding videos from YouTube has become a common strategy for data collection. The YT-ALL dataset \cite{morgado2018self} was curated in the wild by retrieving 360° videos from YouTube using spatial audio-related queries. It is further categorized into two subsets: YT-MUSIC, which consists of live music performance videos, and YT-CLEAN, which includes scenes with fewer overlapping sound sources.
Building upon this approach, YT-360 \cite{morgado2020learning} collected a large-scale dataset of 360° videos with spatial audio from YouTube, covering a broad range of topics such as musical performances, vlogs, sports, and more. This semantic and contextual diversity is crucial for learning robust audio-visual spatial representations. To evaluate the capacity of audio-visual spatial alignment (AVSA) pre-training for downstream tasks such as semantic segmentation, the authors employ the ResNet101 panoramic FPN model \cite{kirillov2019panoptic} to segment objects and background categories within the dataset.
For more precise and context-aware spatial audio synthesis, the YT-AMBIGEN dataset \cite{kim2025visage} was compiled by filtering YouTube clips that include corresponding FOA recordings and camera orientation metadata. To ensure the presence of semantically recognizable audio events in each clip, the authors applied an AudioSet-based classification model \cite{koutini2021efficient} to select segments with high-probability event labels recursively. Visual cues for audio generation were constrained within the field of view (FoV) by computing the audio energy map of each clip, identifying the most likely sound event location (i.e., the peak of the energy map), and cropping the panoramic video around that point to obtain the FoV-aligned video segment.

Beyond data crawling, simulating first-order ambisonics (FOA) is a prevalent strategy for generating spatial audio datasets. The TUT Sound Events 2018 series \cite{adavanne2018sound} includes several representative examples. For instance, the ANSYN dataset synthesizes anechoic scenes using artificial impulse responses and external event samples (e.g., speech, coughing, door slam) \cite{lafay2017sound}, which are randomly positioned, captured via microphone arrays, and converted to FOA. RESYN follows a similar pipeline but simulates reverberant environments using the image-source method. In contrast, the REAL dataset records real acoustic conditions by capturing impulse responses in a university corridor and convolving them with UrbanSound8K events \cite{arnault2020urban}, producing realistic FOA representations. These datasets provide valuable benchmarks for sound event localization and detection (SELD) tasks. 
The DCASE2021 Task 3 dataset \cite{politis2021dataset} simulates spatialized recordings of static and dynamic sound events using real room impulse responses, incorporating more challenging polyphonic scenes and unlabeled directional interference sources. It features events drawn from NIGENS\cite{trowitzsch2019nigens} and is constructed from high-resolution SRIRs.
Building on LibriSpeech \cite{panayotov2015librispeech}, the Spatial LibriSpeech dataset \cite{sarabia2023spatial} augments speech with source positions, orientations, and room acoustics via simulated environments, supporting tasks like 3D localization and reverberation analysis. More recently, SonicSim \cite{li2024sonicsim}, a high-fidelity acoustic simulation toolkit based on Habitat-Sim \cite{savva2019habitat}, enables efficient generation of realistic synthetic audio. Leveraging this framework, SonicSet was developed as a large-scale dataset featuring environments from Matterport3D \cite{chang2017matterport3d}, speech from LibriSpeech, and background sounds from FSD50K \cite{fonseca2021fsd50k} and FMA \cite{defferrard2016fma}; by simulating complex acoustic phenomena such as reflection and diffraction, SonicSet offers high-quality FOA audio with strong spatial realism.

In contrast, FOA (First-Order Ambisonics) data captured in real-world settings is particularly valuable due to its scarcity and realism. The REC-STREET dataset \cite{morgado2018self} was recorded using a 360° camera equipped with a spatial audio microphone and consists of videos depicting outdoor street scenes. Owing to the consistency in recording hardware and scene content, the audio in REC-STREET videos is relatively easier to spatialize, making the dataset well-suited for research on outdoor audio-visual scene understanding.
The STARSS22 dataset \cite{politis2022starss22} comprises spatial recordings collected in diverse indoor environments at two separate locations. Captured using high-resolution spherical microphone arrays, the recordings are provided in two 4-channel spatial formats: First-Order Ambisonics and tetrahedral microphone array format. It includes sound events belonging to 13 target classes, annotated both temporally and spatially through a combination of manual labeling and optical tracking.
Building upon its predecessor, the STARSS23 dataset \cite{shimada2023starss23} further enhances data diversity, modality, and annotation richness. It offers multichannel audio recorded with microphone arrays, synchronized video data, and detailed spatio-temporal annotations of sound events. The acoustic scenes in STARSS23 were deliberately designed and recorded according to predefined instructions, ensuring sufficient occurrence and activity of the target sound events. STARSS23 also provides manually verified temporal activation labels and direction-of-arrival (DoA) annotations, the latter derived from motion capture system tracking results. Our benchmark evaluations demonstrate that incorporating visual object locations significantly benefits audio-visual SELD tasks.

\subsubsection{Binaural Datasets}
Binaural audio provides a perceptually grounded and spatially immersive listening experience by capturing how sound reaches the human ears from different directions, relying on interaural time differences (ITD), interaural level differences (ILD), and spectral filtering introduced by the listener’s head and body, collectively modeled via the Head-Related Transfer Function (HRTF). Compared to microphone arrays and FOA formats, binaural audio more directly emulates human hearing, and when rendered through headphones, it offers highly realistic spatial cues. This has made it increasingly popular in areas such as virtual and augmented reality, immersive media, human-computer interaction, and multimodal audio-visual learning. With the rise of spatially aware audio applications, numerous datasets have been developed to support tasks like speech enhancement, sound event localization and detection (SELD), and spatial audio generation.

A core subset of these datasets consists of real-world recordings using binaural microphones. One of the most notable examples is FAIR-Play \cite{gao20192}, which captures audiovisual clips in a large music room using a binaural microphone and a GoPro camera. The goal is to simulate realistic acoustic scenes through varying combinations of instruments and performers, enabling models to decode monaural input into spatialized binaural output with the help of visual context. The recorded scenes include instruments such as cello, guitar, harp, piano, trumpet, and more, offering rich spatial diversity. Similarly grounded in naturalistic data, the dataset proposed by Richard et al. \cite{richard2021neural} involves eight participants (balanced by gender) recorded while engaging in casual conversation. A GRAS KEMAR dummy head fitted with binaural microphones simulates the listener, while an OptiTrack system tracks the speaker's head pose. This setup enables the generation of spatially accurate, visually guided binaural sound and has been widely adopted in audio-visual spatialization research. Another real-world dataset, EasyCom \cite{donley2021easycom}, was recorded in a noisy, restaurant-like room where participants conversed naturally around a table while ten loudspeakers played background noise. Multimodal recordings include AR glasses with microphone arrays, RGB video, headset audio, annotated speech activity, source locations, and identities, making it particularly valuable for robust modeling in challenging acoustic environments.

Beyond physically recorded data, simulated datasets have enabled researchers to generate large-scale, diverse, and controllable binaural scenes. SimBinaural \cite{garg2023visually} exemplifies this approach by combining SoundSpaces audio simulation \cite{chen2020soundspaces} with the Habitat simulator \cite{savva2019habitat} to synthesize binaural audio within realistic 3D environments from Matterport3D \cite{chang2017matterport3d}. Dry audio signals, such as music or speech from Freesound \cite{font2013freesound} and OpenAIR \cite{murphy2010openair}, are convolved with virtual RIRs to simulate spatial characteristics while avoiding contamination from pre-existing reverberation. The virtual microphone-camera rig follows controlled trajectories, ensuring spatial and visual diversity in each clip. As a complement to this synthetic paradigm, YouTube-Binaural \cite{garg2023visually} augments an existing 360-degree video dataset \cite{morgado2018self} by converting ambisonic audio into pseudo-binaural signals and extracting standard field-of-view clips oriented toward the main sound source. This hybrid approach merges real-world video with synthesized spatial audio to increase realism and scene variability.

Recent datasets have gone further by integrating multimodal and semantic data to support language-driven or cross-modal spatial learning. BEWO-1M \cite{sun2024both} is the first large-scale dataset explicitly designed for stereo and binaural spatial audio generation under textual supervision. It contains over one million audio-text and audio-image pairs with annotated spatial descriptions. The dataset is built by filtering and augmenting audio from prior datasets like VGGSound \cite{chen2020vggsound}, removing short or mismatched segments, and applying GPT-4 \cite{achiam2023gpt} to generate rich spatial descriptions. Additional synthetic augmentation ensures a diverse range of source configurations, including single, dual, multi-source, and mixed dynamic/static scenes. This dataset supports controllable spatial generation grounded in language. In a more musically oriented context, BinauralMusic \cite{li2024binauralmusic} contains video clips with binaural recordings across nine instrument categories, captured in various indoor and outdoor settings like malls, hotels, gardens, and fields. It enables tasks such as vision-guided source localization and separation, and supports audio-visual cross-modal learning more generally.

Finally, a unique contribution comes from the MRSDrama dataset \cite{zhang2025isdrama}, the first spatial multimodal corpus built around dramatic narrative performance. It includes binaural audio, video, drama scripts, geometric poses, and GPT-4o-generated textual prompts. Expressive speakers performed translated excerpts from well-known dramatic scripts across three different real-world scenes, using professional-grade recording and tracking equipment. The audio was denoised and aligned, the video was annotated for speaker actions and positions, and scene context and camera pose information were also labeled. MRSDrama currently stands as the largest recorded spatial speech dataset, enabling high-level tasks in spatial storytelling, speech synthesis, and expressive multimodal generation.

\subsection{Evaluation Metrics}

\subsubsection{Evaluation Metrics for Spatial Audio Understanding}

\textbf{SELD.} The SELD task is evaluated using separate metrics for Sound Event Detection (SED) and Direction-of-Arrival (DOA) estimation. For SED, the one-second segment F-score and Error Rate (ER) are commonly used \cite{mesaros2016metrics}.

For DOA estimation, two frame-wise metrics are frequently employed \cite{adavanne2018direction}: \textbf{DOA Error} and \textbf{Frame Recall}. Let \(T\) be the total number of time frames. Denote by \(\mathbf{DOA}_R^t\) the set of reference DOAs at frame \(t\) and by \(\mathbf{DOA}_E^t\) the set of estimated DOAs. Define
\[
D_R^t = \bigl|\mathbf{DOA}_R^t\bigr|,\quad
D_E^t = \bigl|\mathbf{DOA}_E^t\bigr|.
\]

The \textbf{DOA Error} is defined as
\begin{equation}
\mathrm{DOA\ Error}
= \frac{1}{\sum_{t=1}^T D_E^t}
  \sum_{t=1}^T
    \mathrm{Hungarian}\bigl(\mathbf{DOA}_R^t,\mathbf{DOA}_E^t\bigr),
\end{equation}
where \(\mathrm{Hungarian}(\cdot,\cdot)\) denotes the optimal assignment cost computed by the Hungarian algorithm, using as the pairwise cost the central angle between a reference DOA \((\phi_R,\lambda_R)\) and an estimated DOA \((\phi_E,\lambda_E)\):
\begin{equation}
\sigma
= \arccos\Bigl(\sin\lambda_E\,\sin\lambda_R
  + \cos\lambda_E\,\cos\lambda_R\,\cos\lvert\phi_R - \phi_E\rvert\Bigr).
\end{equation}
Here \(\phi\in[-\pi,\pi]\) is the azimuth and \(\lambda\in[-\tfrac{\pi}{2},\tfrac{\pi}{2}]\) is the elevation.

To account for frames where the number of estimated DOAs does not match the number of reference DOAs, the \textbf{Frame Recall} is defined as
\begin{equation}
\mathrm{Frame\ Recall}
= \frac{1}{T}\sum_{t=1}^T \mathbf{1}\bigl(D_R^t = D_E^t\bigr),
\end{equation}
where \(\mathbf{1}(\cdot)\) is the indicator function, equal to 1 if its argument is true and 0 otherwise.

An ideal SELD method achieves an error rate of zero, an F-score of 1 (100\%), a DOA Error of 0°, and a Frame Recall of 1 (100\%). To compare submitted methods, each method is ranked individually for all four metrics, and final positions are determined by the cumulative minimum of these ranks.

The four cross-validation folds are treated as a single experiment: metrics are computed only after training and testing all folds. Intermediate measures (insertions, deletions, substitutions) are aggregated across folds before calculating the final metrics, rather than averaging per fold \cite{forman2010apples}.

\paragraph{Spatial Audio Separation.}
Metrics To measure the quality of separation, usually adopt the widely-used mir eval library metrics: Signal-to-Distortion Ratio (SDR) measures both interference and artifacts, Signal-to-Interference-Ratio (SIR) measures interference. Higher values indicate a better degree of separation \cite{ye2024lavss}.

\paragraph{Joint Learning.}
In joint learning, they typically employ two binary‐classification‐based evaluation metrics\cite{morgado2020learning}. \textbf{AVC-Bin} (Audio–Visual Correspondence) determines whether an audio–video clip pair originates from the same video instance. \textbf{AVSA-Bin} (Audio–Visual Spatial Alignment) assesses the spatial consistency between the audio and visual streams.

For semantic segmentation, the model’s dense‐prediction capability is evaluated using \textbf{pixel accuracy and mean Intersection-over-Union (mean IoU)}.  Additionally, \textbf{clip-level accuracy} is employed for action recognition.

\subsubsection{Evaluation Metrics for Spatial Audio Generation}

\paragraph{Monaural-to-Binaural Audio Generation.} To comprehensively assess the fidelity of the synthesized binaural signal \(\hat{x}\) concerning the reference binaural recording \(x\), previous works \cite{leng2022binauralgrad,liu2022dopplerbas} on \textbf{monaural-to-binaural audio generation} adopt both objective and subjective criteria. Except for the perceptual measures, PESQ and MOS, all metrics are lower-is-better.  Notation is unified as follows: \(n\in\{1,\dots,T\}\) indicates time-domain samples; \(c\in\{L,R\}\) indexes the two output channels; \(k\in\{1,\dots,K\}\) and \(m\in\{1,\dots,M\}\) denote STFT frequency and frame indices; \(\operatorname{STFT}\{\cdot\}\) yields the complex time–frequency representation.

For \textbf{Wave \(\mathbf{L_2}\)},
The time-domain mean-squared error (MSE) captures sample-by-sample deviations:
\begin{equation}
\begin{aligned}
&\mathcal{L}^{\text{wave}}_{L_2}
  \ =\ 
  \frac{1}{T}
  \sum_{n=1}^{T}
  \sum_{c\in\{L,R\}}
  \bigl(\hat{x}_{c}[n]-x_{c}[n]\bigr)^{2}.
\end{aligned}
\end{equation}
Although it provides a well-behaved gradient and is easy to implement, it ignores the non-uniform frequency sensitivity of human hearing.

For \textbf{Amplitude \(\mathbf{L_2}\)},
after converting both signals to their magnitude spectra,
\begin{equation}
\begin{aligned}
    &X(k,m)=|\operatorname{STFT}\{x\}(k,m)|,\\
&\hat{X}(k,m)=|\operatorname{STFT}\{\hat{x}\}(k,m)|.
\end{aligned}
\end{equation}
The energy envelope mismatch is quantified as
\begin{equation}
\begin{aligned}
&\mathcal{L}^{\text{amp}}_{L_2}
  \ =\ 
  \frac{1}{K M}
  \sum_{k=1}^{K}\sum_{m=1}^{M}
  \bigl(\hat{X}(k,m)-X(k,m)\bigr)^{2}.
\end{aligned}
\end{equation}

For \textbf{Phase \(\mathbf{L_2}\)}, spatial cues rely strongly on interaural phase differences.  To prevent phase-wrap artefacts, we minimize the wrapped phase distance:
\begin{equation}
\begin{aligned}
\mathcal{L}^{\text{phase}}_{L_2}
  \ =&\ 
  \frac{1}{K M}
  \sum_{k=1}^{K}\sum_{m=1}^{M}
  \Bigl(
    \operatorname{wrap}\, \bigl(
      \angle\hat{X}(k,m)\\
      &-\angle X(k,m)
    \bigr)
  \Bigr)^{2}, 
\end{aligned}
\end{equation}
where \(\operatorname{wrap}(\theta)\in[-\pi,\pi)\).

To align perceptual quality with spectral accuracy, we average three complementary losses over a bank of \(M\) STFT configurations \(\{\cdot^{(i)}\}_{i=1}^{M}\) as \textbf{Multi-Resolution STFT Loss (MRSTFT)}:
\begin{equation}
\begin{aligned}
  \mathcal{L}_{\mathrm{SC}}^{(i)}
    &= \frac{\bigl\lVert\,|X^{(i)}| - |\hat X^{(i)}|\,\bigr\rVert_{F}}
           {\bigl\lVert\,|X^{(i)}|\,\bigr\rVert_{F}}, 
  \\[4pt]
  \mathcal{L}_{\mathrm{mag}}^{(i)}
    &= \frac{1}{N^{(i)}}\,\bigl\lVert\,|X^{(i)}| - |\hat X^{(i)}|\,\bigr\rVert_{1}, 
  \\[4pt]
  \mathcal{L}_{\mathrm{log}}^{(i)}
    &= \frac{1}{N^{(i)}}\,\bigl\lVert\,
       \log\bigl(|X^{(i)}| + \varepsilon\bigr)
       - \log\bigl(|\hat X^{(i)}| + \varepsilon\bigr)
      \,\bigr\rVert_{1}.
\end{aligned}
\end{equation}
\begin{equation}
\begin{aligned}
\mathcal{L}_{\text{MRSTFT}}
  \ = \ 
  \frac{1}{M}\sum_{i=1}^{M}
  \bigl(
    \mathcal{L}_{\text{SC}}^{(i)}
    +\lambda_{\text{mag}}\mathcal{L}_{\text{mag}}^{(i)}
    +\lambda_{\text{log}}\mathcal{L}_{\text{log}}^{(i)}
  \bigr).
\end{aligned}
\end{equation}

This compound objective balances global spectral convergence with fine-grained magnitude fidelity across multiple time–frequency resolutions.

For \textbf{Perceptual Evaluation of Speech Quality (PESQ)},
the ITU-T P.862 standard maps symmetric (\(d_{\text{sym}}\)) and asymmetric (\(d_{\text{asym}}\)) perceptual distortions onto a MOS-like scale:
\begin{equation}
\begin{aligned}
&\text{PESQ}
  \ =\ 
  4.5
  \ -\ 0.1\,d_{\text{sym}}
  \ -\ 0.0309\,d_{\text{asym}},
\end{aligned}
\end{equation}
yielding scores in \([-0.5,\,4.5]\).  Higher values denote closer perceptual similarity.

Finally, subjective quality \textbf{Mean Opinion Score (MOS)} is obtained by averaging listener ratings over a five-point Likert scale:
\begin{equation}
\begin{aligned}
&\text{MOS}
  \ = \ 
  \frac{1}{N}\sum_{i=1}^{N}s_{i},
\end{aligned}
\end{equation}
where \(s_{i}\) is the score from the \(i\)-th participant.  MOS serves as the definitive benchmark to which all objective metrics are ultimately calibrated.

Wave/Amplitude/Phase \(L_2\) losses provide gradient-friendly objectives that capture complementary signal aspects. MRSTFT augments them with multi-resolution spectral consistency. PESQ offers a single-ended perceptual estimate that correlates well with telecommunication speech quality, and MOS delivers the gold-standard human judgment.  Together, this metric suite affords a balanced evaluation of both technical accuracy and perceptual realism in mono-to-stereo binaural conversion.

\paragraph{End-to-End Binaural Audio Generation.} Evaluation metrics are highly varied for this task. In the case of binaural spatial audio, metrics can be computed based on interaural time difference (ITD), interaural level difference (ILD), and embeddings from a pretrained spatial-audio-understanding model\cite{zheng2024bat} to calculate specific performance indicators\cite{zhang2025isdrama}. 
For the objective evaluation of IPD and ILD, they first convert the time-domain signal $x(n)$ into the frequency-domain signal $X(t,f)$ using the short-time Fourier transform (STFT):
\begin{equation}
\begin{aligned}
&X_i(t,f) = \sum_{n=0}^{N-1} x_i(n) \cdot w(t-n) \cdot e^{-j2\pi fn}, i \in \{1, 2\},
\end{aligned}
\end{equation}
where \( w(t-n) \) is a window function, \( N \) is the window length, and $i$ indicates the channel of the binaural audio.
Next, they calculate the mel-spectrogram, IPD, and ILD based on the frequency-domain signals $X_i(t,f)$. 
The mel-spectrogram for each channel is calculated as:
\begin{equation}\label{mel-eq}
\begin{aligned}
&S_i(t, m) = \log\left(|X_i(t, f)|^2 \times \text{melW} \right),
\end{aligned}
\end{equation}
where melW is an $M$-bin mel filter bank.
\textbf{IPD} is derived from the phase spectrograms of the left and right channels:
\begin{equation}\label{ipd-eq}
\begin{aligned}
&IPD(t, f)=\angle \frac{X_2(t, f)}{X_1(t, f)}.
\end{aligned}
\end{equation}
Then, \textbf{ILD} is extracted from the loudness spectrum of the left and right channels:
\begin{equation}\label{ild-eq}
\begin{aligned}
& ILD(t, f)=20\log_{10}\left( \dfrac{|X_2(t,f)| + \varepsilon}{|X_1(t,f)| + \varepsilon} \right), \varepsilon = 1e^{-10}.
\end{aligned}
\end{equation}
They calculate Mean Absolute Error (MAE) metrics based on the IPD and ILD extracted from the ground truth (GT) and the predicted speech. 
Since the IPD here is in radians and the ILD uses log10, the resulting values are quite small, especially after averaging the MAE over the time dimension. 
So, they multiply by 100 to make the results more intuitive.

Additionally, they analyze angular and distance metrics using SPATIAL-AST \citep{zheng2024bat}. 
SPATIAL-AST encodes \textbf{angle and distance} embedding for binaural audio. 
They compute and average the cosine similarity for each 1-second segment based on the GT and predicted audio.

\paragraph{End-to-End FOA Generation.}
Current methods usually assess spatial localization accuracy by measuring azimuth error, elevation error, distance error, and spatial‐angle difference \cite{heydari2025immersediffusion}. Codec quality is evaluated via STFT and Mel distances between original and reconstructed FOA audio on the test set, using AuraLoss with default settings \cite{evans2024fast,evans2024long}. Plausibility of generated clips is quantified by the \textbf{Fréchet Audio Distance (FAD)} between generated and reference embeddings, and by \textbf{KL divergence} computed with a pretrained ELSA model. The \textbf{CLAP score}, the cosine similarity between spatial text embeddings and corresponding audio embeddings, is also reported. For the parametric model, KL divergence and CLAP are computed using spatial captions from the test set, despite training on non‐spatial captions and parameters.

To measure spatial accuracy, they compare ground‐truth and estimated \textbf{azimuth \(\theta\), elevation \(\phi\), and distance \(d\)}. Intensity vectors \(I_x, I_y, I_z\) are obtained by multiplying the omnidirectional channel \(W\) with the directional channels \(X, Y, Z\):
\begin{equation}
I_x = W \cdot X,\quad
I_y = W \cdot Y,\quad
I_z = W \cdot Z
\label{eq:intensity}
\end{equation}
\begin{equation}
\theta = \tan^{-1}\frac{I_y}{I_x},\quad
\phi = \tan^{-1}\frac{I_z}{\sqrt{I_x^2 + I_y^2}},\quad
d = \sqrt{I_x^2 + I_y^2 + I_z^2}
\label{eq:combined}
\end{equation}

They report the L1 norm of the differences for azimuth, elevation, and distance. For azimuth, they use the circular difference:
\begin{equation}
\mathrm{L1}_{\theta}
= ||(|\theta - \hat{\theta}|,2\pi - |\theta - \hat{\theta}|\bigr)||_1
\label{eq:circular_difference}
\end{equation}

Spatial‐angle error \(\Delta_{\mathrm{Spatial-Angle}}\) is defined as \cite{van2012heavenly}:
\begin{equation}
a = \sin^2\bigl(\tfrac{\Delta_{\phi}}{2}\bigr)
+ \cos(\phi)\cos(\hat{\phi})\sin^2\bigl(\tfrac{\Delta_{\theta}}{2}\bigr)
\label{eq:alpha}
\end{equation}
\begin{equation}
\Delta_{\mathrm{Spatial-Angle}}
= 2 \arctan2\bigl(\sqrt{a},\,\sqrt{1 - a}\bigr)
\label{eq:angular_distance}
\end{equation}
Here, \(\Delta_{\phi}\) and \(\Delta_{\theta}\) denote the linear and circular differences for elevation and azimuth, respectively.

\section{Conclusion}
\label{sec: con}
In response to the rapid advancements in immersive audio, this paper has presented a comprehensive and systematic survey of the spatial audio field, addressing a notable gap in existing literature. 
Our review provides a structured overview of the domain by organizing recent works according to several key aspects. 
We begin by detailing the various input and output representations that form the foundation of spatial audio tasks. 
The core of our survey then focuses on the two primary research paradigms: spatial audio understanding and spatial audio generation.
In the understanding domain, we explore the methodologies for spatial audio analysis, including sound event detection, localization, sound source separation and joint learning.
In the generation domain, we examine techniques for synthesizing spatial audio, and summarize generative models by their architectures.
To complete our analysis, we systematically summarize the landscape of relevant datasets, evaluation metrics, and benchmarks from both training and assessment perspectives.
We hope that this survey will serve as a valuable resource for researchers and practitioners in the field, guiding future research directions and fostering advancements in spatial audio technologies.

\newpage
\bibliographystyle{neurips_2024}


\end{document}